\documentclass[%
 reprint,
superscriptaddress,
nobibnotes,
 amsmath,amssymb,
 aps,
 prl,
]{revtex4-2}

\usepackage{graphicx}
\usepackage{dcolumn}
\usepackage{bm}
\usepackage{bbm}
\usepackage[colorlinks=true, allcolors=blue]{hyperref}
\usepackage{braket}
\usepackage{mathtools}
\usepackage{stmaryrd}
\usepackage[ruled,linesnumbered,vlined,titlenotnumbered]{algorithm2e} 
\usepackage{tikz}
\usepackage{tikz-cd}
\usepackage{amsthm}

\newtheorem*{maintheorem}{\bf Theorem}

\theoremstyle{definition}

\theoremstyle{remark}

\renewcommand{\tilde}{\widetilde}

\renewcommand{\geq}{\geqslant}

\newcommand{\dkap}{\delta\kern-1.25pt\varkappa}

\newcommand{\GL}{\operatorname{GL}}

\newcommand{\ann}{\operatorname{ann}}
\newcommand{\coker}{\operatorname{coker}}

\newcommand{\im}{\operatorname{im}}
\newcommand{\Ext}{\operatorname{Ext}}

\newcommand{\Hom}{\operatorname{Hom}}

\newcommand{\diag}{\operatorname{diag}}
\newcommand{\id}{\mathrm{id}}

\makeatletter
\newcommand{\RemoveAlgoNumber}{\renewcommand{\fnum@algocf}{\AlCapSty{\AlCapFnt\algorithmcfname}}}
\newcommand{\RevertAlgoNumber}{\algocf@resetfnum}
\makeatother

\begin{document}

\title{Decoupling 2D translation-invariant topological CSS codes}

\author{Yifei Wang}
\affiliation{Institute for Advanced Study, Tsinghua University, Beijing, 100084, China}
\affiliation{Beijing Key Laboratory of Cold Atom Quantum Computation, Tsinghua University, Beijing, 100084, China}

\author{Zhongyi Ni}
\affiliation{Hong Kong University of Science and Technology (Guangzhou), Guangzhou, 511453, China}

\author{Mingxin He}
\affiliation{Institute for Advanced Study, Tsinghua University, Beijing, 100084, China}
\affiliation{Beijing Key Laboratory of Cold Atom Quantum Computation, Tsinghua University, Beijing, 100084, China}

\author{Jinguo Liu}
\affiliation{Hong Kong University of Science and Technology (Guangzhou), Guangzhou, 511453, China}

\author{Yingfei Gu}
\email[E-mail: ]{guyingfei@tsinghua.edu.cn}
\affiliation{Institute for Advanced Study, Tsinghua University, Beijing, 100084, China}
\affiliation{Beijing Key Laboratory of Cold Atom Quantum Computation, Tsinghua University, Beijing, 100084, China}

\date{August 8, 2026}

\begin{abstract}
Two-dimensional translation-invariant topological CSS codes on qubits are known to be locally equivalent, after coarse-graining, to stacks of toric codes. However, existing constructions generally break more translation symmetry than is required to remove anyon-permuting translations, leaving open whether any further obstruction exists. We prove that no such obstruction occurs: after passing to the maximal anyon-preserving superlattice, every such code admits a local unitary decoupling into toric codes and product states. We further provide an efficient algorithm for explicitly constructing the decoupling map, together with bounds on the required supercell size and operator spreading. The decoupling requires no additional ancillas in generic cases and extends to finite systems with suitable boundary conditions.

\end{abstract}

\maketitle

\emph{Introduction.--}
Two-dimensional (2D) topological Calderbank-Shor-Steane (CSS) codes, exemplified by the toric code~\cite{kitaevFaulttolerantQuantumComputation2003}, are central to fault-tolerant quantum computation and the study of topological phases of matter. Since its introduction, the toric code has become one of the best-understood quantum codes, with major theoretical and experimental advances, including the characterization of its error-correction threshold and the development of efficient  decoders~\cite{dennisTopologicalQuantumMemory2002}, fault-tolerant logical gate protocols based on braiding and lattice surgery~\cite{fowlerSurfaceCodesPractical2012,horsmanSurfaceCodeQuantum2012},
and the 
experimental realization of quantum error correction below threshold~\cite{acharyaQuantumErrorCorrection2025}. 

More recently, bivariate bicycle (BB) codes have emerged as a promising family of quantum LDPC codes with a 2D translation structure~\cite{bravyiHighthresholdLowoverheadFaulttolerant2024}. Their relatively high code rates make them attractive candidates for low-overhead quantum memories. However, important questions concerning logical gates and efficient decoding remain active areas of research.  
A deeper understanding of BB codes, and more broadly of 2D translation-invariant topological CSS codes, is therefore important both fundamentally and for fault-tolerant quantum computation. 

A natural strategy is to relate these codes to well-understood models such as the toric code. Such connections allow structural properties and computational tools to be transferred from familiar codes to less understood ones. This approach has been particularly fruitful for color codes, which can be mapped by local unitaries with ancillas to decoupled copies of the toric code~\cite{kubicaUnfoldingColorCode2015}. Our goal is to construct such a decoupling map for a generic 2D translation-invariant topological CSS code.

Existing classification and constructive results establish that these codes are locally equivalent, after suitable coarse-graining, to stacks of toric codes and product states~\cite{bombinUniversalTopologicalPhase2012, bombinStructure2DTopological2014,haahAlgebraicMethodsQuantum2017, haahClassificationTranslationInvariant2021}. 
However, these constructions generally coarse-grain beyond what is required to eliminate anyon-permuting translations, leaving open whether an additional fundamental obstruction exists to decoupling at the maximal residual translation symmetry. 
We show that no such obstruction exists: anyon permutation is the sole obstruction~\footnote{Our results imply that this code is in the same symmetry-enriched topological phase as the decoupled toric codes, where the relevant symmetry is the residue translation symmetry. We acknowledge Liujun Zou for pointing out this perspective.}.
Thus, {\it every 2D translation-invariant topological CSS code admits a decoupling unitary preserving the maximal anyon-preserving translation symmetry.} 
In addition, we explicitly construct an efficient algorithm computing the decoupling unitary with controlled operator spreading.

\emph{Definitions and notation.--}
We consider 2D translation-invariant topological CSS codes defined as the ground-state spaces of commuting Pauli Hamiltonians 
\begin{equation}\label{eq-main: hamiltonian}
    H = -\sum_{(a,b)\in \mathbb{Z}^2}\left(\sum_{i=1}^s S^X_i(a,b) + \sum_{i=1}^r S^Z_i(a,b)\right),
\end{equation}
subject to the following conditions:
\begin{enumerate}
    \item Local CSS stabilizer generators: Each $S_i^{X/Z}(a,b)$ is a finite-support product consisting exclusively of Pauli $X$ or $Z$ operators.
    All generators commute and together generate the stabilizer group. 
    \item Translation invariance: Each  $S_i^{X/Z}(a,b)$ is obtained by translating $S_i^{X/Z}(0,0)$ by lattice vector $(a,b)$.
    \item  Topological order: 
There are no finite-support logical Pauli operators. Equivalently, on the infinite plane, every finite-support Pauli operator commuting with the stabilizer group belongs to the stabilizer group, up to phase (cf. TQO-1 of Bravyi, Hastings, and Michalakis~\cite{bravyiTopologicalQuantumOrder2010}).
\end{enumerate}

We now formulate these conditions using the module-theoretic framework~\cite{haahCommutingPauliHamiltonians2013} and the chain-complex representation of CSS codes~\cite{freedman2002z2, kitaevFaulttolerantQuantumComputation2003, bravyiQuantumCodesLattice1998}. 
Let $R=\mathbb F_2[x^{\pm1},y^{\pm1}]$ 
be the Laurent polynomial ring, where multiplication by $x$ and $y$ represents translation by one lattice spacing in the two coordinate directions. The code is customarily represented by the chain complex
\begin{equation}
R^r \xrightarrow{\ H_Z^\dagger\ } R^q
\xrightarrow{\ H_X\ } R^s .
\label{eq-main: chain complex}
\end{equation}
where $\dagger$ denotes transposing a matrix and taking element-wise antipodes,
i.e., replacing variables $x,y$ by $\overline{x}=x^{-1}, \overline{y}=y^{-1}$.
Pauli operators are represented modulo overall phases, so their multiplication corresponds to addition in the associated $R$-modules. The middle module $R^q$ represents Pauli $Z$ operators on a lattice with $q$ qubits per unit cell: the element
$x^a y^b e_i\in R^q$
corresponds to the operator $Z_i(a,b)$ acting on qubit $i$ in the unit cell at $(a,b)$.
The module $R^r$ labels the $r$ translation classes of $Z$-type stabilizer generators. In particular, $x^a y^b e_i\in R^r$ is mapped by $H_Z^\dagger$ to the stabilizer $S_i^Z(a,b)$. Similarly, $R^s$ records the syndromes associated with the $s$ translation classes of $X$-type stabilizer generators, and $H_X$ maps a Pauli $Z$ operator to the $X$-check syndrome that it creates. 
The commutativity of the stabilizer generators is equivalent to
$H_XH_Z^\dagger=0$.
The topological-order condition further implies exactness at the qubit module,
$\ker H_X=\operatorname{im}H_Z^\dagger$. The superselection sectors of $X$-check excitations, which we refer to as $e$-anyons, are therefore classified by
$\coker H_X\coloneqq R^s/\im H_X$.
Exchanging the roles of $X$ and $Z$ gives the dual chain complex
$ R^s\xrightarrow{H_X^\dagger} R^q \xrightarrow{H_Z} R^r$ 
whose cokernel $\operatorname{coker}H_Z$ classifies the $m$-anyons.

Without loss of generality, we assume that the chosen translation classes of stabilizer generators are independent, so that both $H_Z^\dagger$ and $H_X^\dagger$ are injective. Such a choice can always be made for two-dimensional topological stabilizer codes
\cite{haahCommutingPauliHamiltonians2013, bombinStructure2DTopological2014}. 
Together with the topological order condition, this implies (1) $q=r+s$, i.e., the density of qubits equals the density of independent checks; and (2) $t:=\dim_{\mathbb{F}_2}\coker H_X = \dim_{\mathbb{F}_2}\coker H_Z<\infty$, i.e., we have equal numbers of $e$ anyons and $m$ anyons.

As a basic example, the toric codes have 
$r = s = 1$ and $q = 2$, with
\begin{equation}
    H_X = (x-1,y-1), \quad H_Z = (1-\overline{y},\overline{x}-1).
    \label{eq-main: toric code}
\end{equation}
Let $J \coloneqq  (x-1, y-1)$.
Then the $e$- and $m$-anyons are both $R/J=\mathbb{F}_2$,
so we have $t = 1$ for toric codes. 

\emph{Translation-symmetry and the decoupling unitary.--}
As briefly discussed in the introduction,
the central question is whether the decoupling map can preserve
the maximal residual translation symmetry allowed by
the code structure.

A certain amount of translation-symmetry breaking is unavoidable because of a phenomenon referred to by Kitaev as weak symmetry breaking~\cite{kitaevAnyonsExactlySolved2006}. In modern terminology, this is also called an anyon-permuting translation symmetry~\cite{barkeshliSymmetryFractionalizationDefects2019}: although the Hamiltonian and ground-state space are translation-invariant, the induced action on superselection sectors can be nontrivial. 
Familiar examples include the Wen plaquette model~\cite{wenQuantumOrdersSymmetric2002, wenQuantumOrdersExact2003}, equivalently the rotated surface code~\cite{bombinOptimalResourcesTopological2007}, and color codes~\cite{bombinTopologicalQuantumDistillation2006,bombinStatisticalMechanicalModels2008}. By contrast, translations act trivially on the anyon types of the standard toric code.

Therefore, the anyons of a generic input code cannot be aligned with those of the target toric codes while preserving the full translation symmetry. One must pass to a superlattice whose translations preserve every anyon, or equivalently coarse-grain the lattice. After this necessary coarse-graining, the code satisfies
\begin{equation}
    \ann\coker H_X
    = \ann\coker H_Z
    = (x-1,\, y-1),
    \label{eq-main: annihilator condition}
\end{equation}
where $\operatorname{ann}M=\{f\in R:fM=0\}$ denotes the annihilator of an $R$-module $M$. Physically, this condition means that every anyon type can be translated by one coarse-grained lattice vector using a finite-support Pauli operator, also referred to as a hopping operator in Ref.~\cite{bombinUniversalTopologicalPhase2012}.

Previous constructions~\cite{bombinUniversalTopologicalPhase2012, bombinStructure2DTopological2014, haahAlgebraicMethodsQuantum2017, haahClassificationTranslationInvariant2021} require further coarse-graining beyond this anyon-preserving superlattice. Is it a genuine obstruction? We show  by the following main theorem that it is not: once the anyon-permuting translations are removed, no further obstruction appears.

\begin{maintheorem}[Symmetry-preserving unitary decoupling]
\label{thm: main}
Given a code whose chain complex and dual complex
\begin{equation}
    R^r\xrightarrow{H_Z^\dagger} R^q \xrightarrow{H_X} R^s,\quad  R^s\xrightarrow{H_X^\dagger } R^q \xrightarrow{H_Z } R^r
\end{equation}
have the following properties: 
\begin{enumerate}
    \item both complexes are exact;
    \item $H_Z^\dagger, H_X^\dagger$ are injective;
    \item $\dim_{\mathbb{F}_2} \coker H_X =\dim_{\mathbb{F}_2}\coker H_Z= t<\infty$,
and  $\ann\coker H_Z = \ann \coker H_X = (x-1,y-1)$,
\end{enumerate}
then there exists a chain isomorphism between the first chain complex and 
a $(t,r,s)$-standard chain complex representing $t$ copies of toric codes, together with 
$p_Z = r - t$ copies of $Z$-basis product states,
and $p_X = s - t$ copies of $X$-basis product states
with check matrices
\begin{equation}
    \begin{aligned}
        \tilde H_X &= \begin{pmatrix}
            I_{p_X} & 0_{p_X\times p_Z} & 0_{p_X\times t} & 0_{p_X\times t} \\
            0_{t\times p_X} & 0_{t\times p_Z} & (x-1) I_t & (y-1)I_t
        \end{pmatrix},\\
        \tilde H_Z^\dagger &= \begin{pmatrix}
            0_{p_X\times p_Z} & 0_{p_X\times t} \\ 
            I_{p_Z} & 0_{p_Z\times t} \\ 
            0_{t\times p_Z} & (1-y)I_{t} \\ 
            0_{t\times p_Z} & (x-1)I_{t} \\ 
        \end{pmatrix}.
    \end{aligned}
    \label{eq-main: goal matrices}
\end{equation}
That is, there exist $R$-isomorphisms $(\psi_2, \psi_1, \psi_0)$ such that the following diagram commutes:
\begin{equation}
\begin{tikzcd}
	{\textnormal{input:}} & {R^{r}} & {R^{q}} & {R^s} \\
	{\textnormal{standard:}} & {R^{r}} & {R^{q}} & {R^s}
	\arrow["{{{H_Z^\dagger}}}", from=1-2, to=1-3]
	\arrow["{{{H_X}}}", from=1-3, to=1-4]
	\arrow["\psi_2", from=1-2, to=2-2]
	\arrow["{{{\tilde{H}_Z^\dagger}}}"', from=2-2, to=2-3]
	\arrow["{\psi_1}", from=1-3, to=2-3]
	\arrow["{{{\tilde{H}_X}}}"', from=2-3, to=2-4]
	\arrow["{{{\psi_0}}}", from=1-4, to=2-4]
\end{tikzcd}
\label{eq-main: chain map}
\end{equation}
The following diagram then automatically commutes for the dual chain:
\begin{equation}
\begin{tikzcd}
	{R^{s}} & {R^{q}} & {R^r} \\
	{R^{s}} & {R^{q}} & {R^r}
	\arrow["{{{{H_X^\dagger}}}}", from=1-1, to=1-2]
	\arrow["{(\psi^\dagger_0)^{-1}}", from=1-1, to=2-1]
	\arrow["{{{{H_Z}}}}", from=1-2, to=1-3]
	\arrow["{{(\psi^\dagger_1)^{-1}}}", from=1-2, to=2-2]
	\arrow["{{{{(\psi^\dagger_2)^{-1}}}}}", from=1-3, to=2-3]
	\arrow["{{{{\tilde{H}_X^\dagger}}}}"', from=2-1, to=2-2]
	\arrow["{{{{\tilde{H}_Z}}}}"', from=2-2, to=2-3]
\end{tikzcd}
\end{equation}
Furthermore, the following algorithm computes the isomorphisms in time polynomial in $q$ and the degrees of the input check matrices $H_Z, H_X$.
\end{maintheorem}

\RemoveAlgoNumber
\begin{algorithm}[htb]
\SetAlgoLined
\KwIn{check matrices $H_Z, H_X$}
\KwOut{invertible maps $(\psi^{-1}_2, \psi^{-1}_1, \psi^{-1}_0)$}
Find 
$\xi_2\in\mathrm{GL}(r,\mathbb F_2)$,
$\xi_1\in\mathrm{GL}(q,\mathbb F_2)$, and
$\xi_0\in\mathrm{GL}(s,\mathbb F_2)$ 
such that $H^\dagger_Z \xi_2 = \xi_1\tilde H_Z^\dagger \pmod J$ and $H_X \xi_1 = \xi_0 \tilde H_X \pmod J$ using standard linear algebra over $\mathbb{F}_2$\;
$H_X \gets \xi_0^{-1}H_X\xi_1$, $H_Z^\dagger\gets \xi_1^{-1}H_Z^\dagger \xi_2$\;
Solve $\tilde{H}_X = H_X \phi_1$ for $(\phi_1)_{q\times q}$, with the columns $[p_X + 1:p_X + p_Z]$ of $\phi_1$ taken from the first $p_Z$ columns of $H_Z^\dagger$\;
Solve  $H^\dagger_Z \phi_2  = \phi_1 \tilde{H}_Z^\dagger$
for $(\phi_2)_{r\times r}$
\;
Replace the lower-right $t\times t$ corner $B$ of $\phi_2$ by $B\bmod J$ to obtain $\phi_2'$\;
Solve $\eta\tilde{H}_Z^\dagger = \phi_2'-\phi_2$
 for $\eta_{r\times q}$ 
\;
$\phi_1' \gets \phi_1 + H_Z^\dagger \eta$\;
\Return $(\xi_2\phi_2', \xi_1 \phi_1', \xi_0)$\;
\caption{Solving for the decoupling unitary\footnote{For technical reasons, the inverses of the isomorphisms $(\psi_2, \psi_1, \psi_0)$ are the natural outputs of our construction.
We do not include a final inversion step in our algorithm,
since matrix inversion is standard in computer algebra and only adds polynomial terms to the time complexity and matrix degree \cite{coxUsingAlgebraicGeometry2005, liang2022degree}.}}
\label{alg-main: constructing isomorphism}
\end{algorithm}

Physically, the isomorphism $\psi_1$ specifies the action of the decoupling unitary on Pauli $Z$ operators, while $(\psi_1^\dagger)^{-1}$ specifies its action on Pauli $X$ operators. The maps $\psi_2$ and $\psi_0$ correspond to changes of basis among the $Z$- and $X$-type stabilizer generators, respectively. 

Before sketching the proof and construction,
we remark why the problem is nontrivial
despite its resemblance to a matrix normal-form problem:

First, the coefficient ring
$R=\mathbb F_2[x^{\pm1},y^{\pm1}]$ is not a field, and Gaussian-style elimination over polynomial rings can suffer from severe intermediate expression swell and therefore exponential running time. 
Previous constructive classifications \cite{haahAlgebraicMethodsQuantum2017, haahClassificationTranslationInvariant2021} control this growth through iterative coarse-graining, thereby breaking more translation symmetry than is necessary.

Second, solving the chain-map equations alone does not generally produce invertible maps. The obstruction is that
$\ker H_X=\im H_Z^\dagger$ is nonzero and in general not a direct summand of the ambient free module $R^q$. 
Consequently, the standard linear algebra techniques available over fields do not apply. Our construction overcomes these two difficulties by combining an initial reduction modulo $J$ with a subsequent correction that restores invertibility.

\emph{Sketch of the proof and construction.--}
The proof proceeds in three steps. First, we reduce the complexes modulo $J=(x-1,y-1)$ to align them with the standard complex, and rewrite them as short exact sequences by removing the cokernels. Second, we interpret these sequences as extensions and compare their extension classes to establish the existence of a chain isomorphism. Third, we correct an explicitly constructed chain map so that each component is invertible.

We begin by reducing the input and standard complexes modulo $J$. Their
homology groups are isomorphic; physically, imposing periodic boundary
conditions on a single coarse-grained unit cell produces equivalent finite
codes. Standard linear algebra over $\mathbb F_2$ therefore yields invertible
matrices $(\xi_2,\xi_1,\xi_0)$ satisfying
\begin{equation}
H_Z^\dagger\xi_2 \equiv
\xi_1\widetilde H_Z^\dagger, \quad H_X\xi_1 \equiv \xi_0\widetilde H_X
\pmod J,
\label{eq-main: clear H mod J}
\end{equation} 
whose role is to identify and separate the product-state sectors $R^{p_{X}}$ from the toric-code sectors $JR^t$.

This transforms our problem to finding isomorphisms $(\phi_2, \phi_1)$ for the following commuting diagram
\begin{equation}
\begin{tikzcd}
	0 & {R^{r}} & {R^{q}} & {R^{p_X}\oplus JR^t} & 0 \\
	0 & {R^{r}} & {R^{q}} & {R^{p_X}\oplus JR^t} & 0
	\arrow[from=1-1, to=1-2]
	\arrow["{H_Z^\dagger}", from=1-2, to=1-3]
	\arrow["{H_X}", from=1-3, to=1-4]
	\arrow[from=1-4, to=1-5]
	\arrow[from=2-1, to=2-2]
	\arrow["\phi_2", from=2-2, to=1-2]
	\arrow["{\tilde{H}_Z^\dagger}"', from=2-2, to=2-3]
	\arrow["{\phi_1}", from=2-3, to=1-3]
	\arrow["{\tilde{H}_X}"', from=2-3, to=2-4]
	\arrow["{\id}", from=2-4, to=1-4]
	\arrow[from=2-4, to=2-5]
\end{tikzcd}
\label{eq-main: ses map}
\end{equation}
with $H_Z = \tilde H_Z \pmod J$ and $H_X = \tilde H_X \pmod J$,
and $\id$ is the isomorphism induced by the chosen generators of the modules.
Now, each row of Eq.~\eqref{eq-main: ses map} defines an extension of
$R^{p_X}\oplus JR^t$ by $R^r$. These extensions are classified by
\begin{equation}
\Ext^1_R\!\left(R^{p_X}\oplus JR^t,R^r\right)
\cong
\Hom_{\mathbb F_2}\!\left(\mathbb F_2^t,\mathbb F_2^r\right).
\end{equation}
The existence of the maps $(\phi_2,\phi_1)$ is therefore equivalent to
the existence of an automorphism $\phi_2$ of $R^r$ such that the pushout
extension
\begin{equation}
0\longrightarrow R^r
\xrightarrow{\,H_Z^\dagger\circ\phi_2\,}
R^q
\xrightarrow{\,H_X\,}
R^{p_X}\oplus JR^t
\longrightarrow 0
\end{equation}
is equivalent to the standard extension.

Under the above identification of the extension group, both the input and
standard extensions are represented by full-rank $r\times t$ matrices over
$\mathbb F_2$. The pushout by $\phi_2$ acts on these representatives by left
multiplication with the invertible matrix $\phi_2^{-1}\bmod J$. Since
$r\geq t$, the group $\mathrm{GL}(r,\mathbb F_2)$ acts transitively on the
set of full-rank $r\times t$ matrices. Hence one can choose $\phi_2$ so that
the input extension is mapped to the standard one. The equivalence of the
two extensions then supplies the middle isomorphism $\phi_1$, proving the
existence of the desired chain isomorphism.

Finally, we turn to the explicit construction. By the comparison theorem for
projective resolutions, the chain maps $(\phi_2,\phi_1)$ are unique up to chain
homotopy. In particular, the last $t$ columns of $\phi_2$ are determined modulo
$J$ and satisfy
\begin{equation}
    (\phi_2)_{[:, p_Z+1:r]} \equiv
    \begin{pmatrix}
        0 \\
        B
    \end{pmatrix} \pmod{J},
\end{equation}
with $B\in\GL(t,\mathbb F_2)$.
Construct an ansatz $\phi_1$ from solving $H_X \phi_1 = \tilde H_X$,
with the columns $[p_X+1:p_X+p_Z]$ in $\phi_1$ taken from the first $p_Z$ columns in $H_Z^\dagger$.
This is possible because the corresponding columns in $\tilde H_X$ are zero (see Eq.~\eqref{eq-main: goal matrices}),
so that the corresponding columns in $\phi_1$ are free to be taken from $\im H_Z^\dagger$.
Solving subsequently $H_Z^\dagger\phi_2=\phi_1\widetilde H_Z^\dagger$
yields a map of the block form
\begin{equation}
\phi_2=
\begin{pmatrix}
I_{p_Z} & C\\
0 & B+D
\end{pmatrix},
\label{eq-main: initial form rho}
\end{equation}
where every entry of $C$ and $D$ lies in $J=(x-1,y-1)$.

We then construct a map $\eta:R^q\to R^r$ satisfying
\begin{equation}
\eta\widetilde H_Z^\dagger
=
\diag(0,-D).
\end{equation}
The existence of such an $\eta$ follows from a direct block-matrix
calculation. The chain-homotopy transformation
\begin{equation}
(\phi_2,\phi_1)
\longmapsto
\bigl(
\phi_2+\eta\widetilde H_Z^\dagger,\,
\phi_1+H_Z^\dagger\eta
\bigr)
\end{equation}
preserves the commutativity of diagram~\eqref{eq-main: ses map}. After this
correction, the lower-right block of $\phi_2$ is $B$, so the corrected
$\phi_2$ is invertible. The Five Lemma then implies that the corrected
$\phi_1$ is also invertible~\cite{weibelIntroductionHomologicalAlgebra1994}.
The algorithm above summarizes this construction; a complete proof is given
in Supplemental Sec.~IV.

We conclude this section with three remarks concerning complexity and locality, ancilla removal, and periodic boundary conditions.
\begin{enumerate}
    \item \emph{Complexity and locality.--}
Our construction avoids the Quillen--Suslin subroutine \cite{quillenProjectiveModulesPolynomial1976, suslinProjectiveModulesPolynomial1976, logarAlgorithmsQuillenSuslinTheorem1992} by exploiting the structure of $H_Z^\dagger$ when choosing the ansatz for $\phi_1$,
yielding an algorithm that only requires solving matrix equations.
This enables us to derive the polynomial behavior of time complexity and operator spreading from results on Gr\"obner basis algorithms \cite{liang2022degree}.

Recall that $q$ denotes the number of qubits in the coarse-grained supercell. 
The locality of the decoupling unitary is controlled by the Laurent-polynomial degrees of $\psi_1$ and $\psi_1^{-1}$, which determine the spreading of local Pauli $Z$ and Pauli $X$ operators, respectively.
We prove in Supplemental Sec.~VII that both the time complexity and the locality are bounded by a polynomial in $q$ and the degrees of the input matrices.
Furthermore, we prove in Supplemental Sec.~III.B that $q < q_02^{q_0^3l^2}$, where $q_0$ is the qubit number per unit cell before coarse-graining, and $l$ is the interaction range.
The locality of the decoupling map can thus be bounded by the initial parameters of the code Hamiltonian.

We test our algorithm
on the BB codes taken from Refs.~\cite{liangGeneralizedToricCodes2025a, zhouBunnyCodesBroadening2026, liangTopologicalSubsystemBivariate2026, symonsSequencesBivariateBicycle2025, bravyiHighthresholdLowoverheadFaulttolerant2024},
and obtain the empirical scalings $T\propto q^{1.86}$
and $\deg(\psi_1^{-1})\propto q^{0.65}$ on these samples.
See Supplemental Sec.~V.C for the detailed numerical results.

\item \emph{On ancilla removal.--}
The decoupling unitary guaranteed by our main theorem
can be compiled to a constant-depth circuit consisting only of CNOT gates.
To see this, we decompose the isomorphic Pauli map as 
$\psi_1 = \diag(\det \psi_1, 1,\ldots, 1)\psi_1'$ with the matrix $\psi_1'$ satisfying $\det \psi_1' = 1$.
For $q\geq 3$, the Suslin stability theorem~\cite{suslinStructureSpecialLinear1977, parkAlgorithmicProofSuslins1994} implies that $\psi_1'$
can be decomposed into elementary matrices over $R$, each corresponding
to translation-invariant layers of CNOT gates. 
The remaining diagonal factor represents a shift (on one qubit per supercell),
since $\det \psi_1$ is a monomial. 
On a product-state sector, this shift acts trivially,
while on a toric-code sector it can also be implemented by a CNOT-only
circuit; see Supplemental Sec.~VI.A.

This conclusion is consistent with the classification of Clifford quantum cellular automata~\cite{haahCliffordQuantumCellular2021}.
A nontrivial QCA index characterizes the transport of local quantum
information, whereas information encoded in a topological code is
necessarily nonlocal. Consequently, a shift may become
circuit-trivial when restricted to the code space.

Finally, the product-state sectors in the standard form need not be
regarded as indispensable ancillas. 
These sectors can be absorbed into the toric-code sectors through the inverse
of the decoupling procedure on toric codes with deliberately enlarged supercell.
Switching between rotated and
unrotated surface-code layouts during syndrome extraction provides a
related example~\cite{mcewenRelaxingHardwareRequirements2023}.

\item \emph{Periodic boundary conditions.--}
Our decoupling extends directly to finite systems with periodic
boundary conditions on the coarse-grained superlattice. For positive integers $L_x$ and $L_y$, imposing the periodic boundary conditions $x^{L_x}=1,
y^{L_y}=1$ corresponds to the base change $R\rightarrow R_{L_x,L_y}:=R/(x^{L_x}-1,y^{L_y}-1)$.
Here, $x$ and $y$ represent translations by one coarse-grained
supercell, so these boundary conditions describe tori whose periods
are integer multiples of the anyon-preserving translation periods. 

As $(\psi_2,\psi_1,\psi_0)$ are $R$-module isomorphisms,
tensoring the input and standard complexes with $R_{L_x,L_y}$
preserves the chain isomorphism. 
Hence the decoupling map descends
directly to the finite torus and maps the code to $t$ copies of the toric code together with the product-state sectors. Since each toric code on a torus encodes $2$ logical qubits, the original finite code encodes $2t$ logical qubits.
In the special case $L_x=L_y=1$, the decoupling then becomes a decoding
transformation that maps the code space onto $2t$ designated physical
qubits tensored with fixed product states.

Alternatively, one may impose periodic boundary conditions directly on the original, uncoarse-grained lattice~\cite{liangGeneralizedToricCodes2025a}. Such boundary conditions need not be compatible with the anyon-preserving superlattice, as occurs for the $\llbracket 144,12,12\rrbracket$ gross code \cite{bravyiHighthresholdLowoverheadFaulttolerant2024}.
This raises the interesting question of how our results can be generalized to such cases.
\end{enumerate}

\emph{Discussions and outlooks.--} 
One potential application is the construction of decoders
for general 2D topological CSS codes.
The measured syndrome can first be mapped to virtual
toric-code sectors, decoded independently using
minimum-weight perfect matching, and then lifted back
to the original code.
The main computational cost is therefore that of graph matching
on a fixed number of coarse-grained toric lattices.
However, the decoupling map spreads a local physical fault
into correlated errors across several virtual sectors.
Independent decoding of these sectors neglects such correlations
and may therefore perform worse than decoders designed directly
for the original code.
This observation also helps explain the advantages of
locality-preserving approaches such as the symmetry decoder
and the cell-matching decoder
\cite{sahay2026matching,tan2026generalized}.

The explicit decoupling map also provides a systematic framework
for studying logical operators and logical gates.
Logical gates implemented by constant-depth circuits in 2D topological stabilizer codes
are restricted to Clifford gates
\cite{bravyiClassificationTopologicallyProtected2013,jochym-oconnorDisjointnessStabilizerCodes2018}.
Known examples include transversal Clifford gates for color codes
and self-dual BB codes
\cite{bombinTopologicalQuantumDistillation2006,liangChenSelfdualBB2025},
as well as fold-transversal gates for surface codes and BB codes
\cite{moussaTransversalCliffordGates2016,
eberhardtLogicalOperatorsFoldtransversal2024}.
By composing these constructions with our decoupling maps,
logical-gate protocols can be transferred between different
2D topological CSS codes with only a bounded loss of locality.
Another class of fault-tolerant constructions is based on
boundaries, defects, and their code deformations
\cite{bravyiQuantumCodesLattice1998, fowlerSurfaceCodesPractical2012}.
A natural direction is to generalize such constructions to BB codes
\cite{eberhardtPruningQLDPCCodes2024,
liangPlanarQuantumLowDensity2025,
liangOperatorAlgebraAlgorithmic2026}
and to characterize how their boundary and defect degrees of freedom
transform under explicit decoupling unitaries.
From a complementary viewpoint, anyon-permuting translations
may themselves act as logical gates requiring no physical operation
beyond relabeling qubits, as explored recently for bivariate
and generalized bicycle codes
\cite{kohEntanglingLogicalQubits2026,
ismailTransversalArchitectureMegaquopscale2026,
davenportGeneralizedBicycleCodes2026}.

Several questions remain open.
For practical applications, it may be advantageous to relax
the requirement of a chain isomorphism to that of a
quasi-isomorphism, potentially yielding maps with smaller spreading
for decoding or logical-gate constructions. 
It is also unclear whether the present method can be extended
to non-CSS stabilizer codes while avoiding the Quillen--Suslin
subroutines and iterative coarse-graining required in previous works.

\emph{Acknowledgment.--}
We thank Yu-An Chen, Ke Liu and Hao Song for discussions and comments on the manuscript. This work is supported by the National Natural Science Foundation of China under Grant No. 12575022, SRICSPYF-ZY2025157, the Shanghai Committee of Science and Technology under Grant No. 25LZ2600800, and the Tsinghua Dushi Program.

\emph{Note added.---} Near the completion of this work, we became aware of independent work by Hao Song~\cite{SongNew}. By introducing a classification method based on Schanuel's lemma---rather than providing a constructive realization of the decoupling unitary---Song establishes the same existence result for two-dimensional decoupling. He also presents an explicit counterexample ruling out a naive three-dimensional generalization and identifies a viable route toward higher-dimensional extensions. We coordinated the timing of our arXiv postings and thank Hao Song for sharing these results prior to their public release.

\bibliography{main}

\end{document}


\title{Supplemental Material: Decoupling 2D translation-invariant topological CSS codes}

\author{Yifei Wang}
\affiliation{Institute for Advanced Study, Tsinghua University, Beijing, 100084, China}
\affiliation{Beijing Key Laboratory of Cold Atom Quantum Computation, Tsinghua University, Beijing, 100084, China}

\author{Zhongyi Ni}
\affiliation{Hong Kong University of Science and Technology (Guangzhou), Guangzhou, 511453, China}

\author{Mingxin He}
\affiliation{Institute for Advanced Study, Tsinghua University, Beijing, 100084, China}
\affiliation{Beijing Key Laboratory of Cold Atom Quantum Computation, Tsinghua University, Beijing, 100084, China}

\author{Jinguo Liu}
\affiliation{Hong Kong University of Science and Technology (Guangzhou), Guangzhou, 511453, China}

\author{Yingfei Gu}
\email[E-mail: ]{guyingfei@tsinghua.edu.cn}
\affiliation{Institute for Advanced Study, Tsinghua University, Beijing, 100084, China}
\affiliation{Beijing Key Laboratory of Cold Atom Quantum Computation, Tsinghua University, Beijing, 100084, China}

\date{August 8, 2026}

\maketitle

\onecolumngrid

\tableofcontents

\section{Module-theoretic language for translation-invariant codes}
\label{sec: fundamental}
We begin by establishing notation and reviewing the module-theoretic framework for translation-invariant topological CSS codes (see \cite{haahAlgebraicMethodsQuantum2017} for lecture notes on the subject).
This defines the object we study in this work
and the properties we take as our starting point.
This section is intended to be pedagogical.
Readers familiar with this language may refer directly to Eq.~(\ref{eq: code chain}) for our convention.

\paragraph{Symplectic formalism.}
To build intuition, we first review a more commonly used construction: representing the Pauli group on a finite number of qubits as a vector space and using a symplectic form to encode the commutation relations.
See Ref.~\cite{kitaevClassicalQuantumComputation2002} for a more complete discussion.
The Pauli group $\mathcal{P}_q$ on $q$ qubits is generated by the single-qubit Pauli operators together with the phase factors $\pm 1, \pm\mathrm{i}$.
The quotient of $\mathcal{P}_q$ by its phase center $\langle \mathrm{i} I \rangle$ is isomorphic, as an additive group, to $\mathbb{F}^{2q}_2$.
Explicitly, fixing a canonical basis $\{e_1,\dots,e_q;\, f_1,\dots,f_q\}$ for this space,
we define the surjective map $\pi\colon \mathcal{P}_q \to \mathbb{F}_2^{2q}$ by $\pi(X_i) = e_i$ and $\pi(Z_i) = f_i$,
extended linearly so that operator multiplication corresponds to vector addition.
We then equip the space with the symplectic form $\omega(e_i,f_j) = \delta_{ij}$.
In matrix form, we denote $\omega(u,v) = u^\top \omega v$, with 
\begin{equation}
    \omega = \begin{pmatrix}
        0 & I_q \\ -I_q & 0
    \end{pmatrix}.
\end{equation}
This symplectic form encodes the commutation relations as $P_1 P_2 = (-1)^{\omega(\pi(P_1),\pi(P_2))}\, P_2 P_1$.

The normalizer of the Pauli group is called the Clifford group,
i.e., given the Pauli group $\mathcal{P}_q$ on $q$ qubits,
the Clifford group is the subgroup of unitary operators satisfying $U\mathcal{P}_q U^\dagger \subseteq \mathcal{P}_q$.
It can be generated by the single-qubit Hadamard and $\pi/2$-phase gates $H, S$, together with the two-qubit CNOT gate.
Clifford operators can be represented as symplectic transformations in the following way:
if $u \in \mathbb{F}_2^{2q}$ represents a Pauli operator $P$
and $U$ is a Clifford unitary, then there exists a unique
$2q\times 2q$ $\mathbb{F}_2$-matrix $V$
such that $U P U^\dagger$ is represented by $V u$.
Since conjugating by unitary preserves commutation relations,
$V$ must be symplectic, i.e., $V^\top \omega V = \omega$.

\paragraph{Lattice system.}
In Ref.~\cite{haahCommutingPauliHamiltonians2013},
Haah introduces the module theoretical language
to generalize the symplectic formalism to lattice systems.
Concretely, for 2D lattices,
we begin with a unit cell containing $q$ qubits.
The lattice is generated by translating this unit cell along two directions $x$ and $y$.
The Pauli group on the full lattice is likewise generated by translates of the unit-cell Pauli group $\mathcal{P}_q$;
we denote the translation of a unit-cell Pauli operator $P$ by a lattice vector $(a,b)\in \mathbb{Z}^2$ as $T(a,b)\, P$.
To extend the representation, we enrich the standard basis $\{e_1,\dots,e_q;\, f_1,\dots,f_q\}$ by introducing monomial prefactors
and define the surjective map by $\pi(T(a,b)\,X_i) = x^a y^b\, e_i$ and $\pi(T(a,b)\,Z_i) = x^a y^b\, f_i$, extended by linearity.
In module-theoretic terms, we have introduced a free module over the Laurent polynomial ring $R = \mathbb{F}_2[x^{\pm 1}, y^{\pm 1}]$,
generated by $\{e_1,\dots,e_q;\, f_1,\dots,f_q\}$,
together with a surjective map satisfying $\pi(T(a,b)\,P) = x^a y^b\, \pi(P)$ for every Pauli operator $P$ on the lattice.

Similarly, in this setup, we can introduce a sesquilinear symplectic form to encode the commutation relations.
The form is defined in matrix notation by
\begin{equation}
    R \ni \omega(u,v) = u^\dagger \omega v = \begin{pmatrix}
        u_1(x^{-1},y^{-1}) & \ldots &u_{2q}(x^{-1},y^{-1})
    \end{pmatrix}\begin{pmatrix}
        0 & I_q \\ -I_q & 0
    \end{pmatrix}\begin{pmatrix}
        v_1(x,y) \\  \vdots  \\ v_{2q}(x,y)
    \end{pmatrix}, \quad \forall u,v\in R^{2q}.
    \label{eq: module symplectic def}
\end{equation}
Here a module element $u$ is written as a column vector, 
and its involution, denoted by $u^\dagger$,
is defined by transposition \textit{and} taking the antipodes of the variables.
In what follows, we will write $\bar{x} \coloneqq x^{-1}$ and $\bar{y} \coloneqq y^{-1}$ for notational brevity.
The commutation relations of Pauli operators can be recovered from this form in the following sense:
$P_1 P_2 = (-1)^{[\omega(\pi(P_1),\pi(P_2))]_0}\, P_2 P_1$,
where $[\,\cdot\,]_0$ denotes the constant term (i.e., the coefficient of $x^0 y^0$).
In fact, the full polynomial $\omega(\pi(P_1),\pi(P_2))$ simultaneously encodes the commutation relations between all translates of $P_1$ and $P_2$:
the coefficient of $x^{a-c} y^{b-d}$ gives the commutation relation between $T(a,b)\,P_1$ and $T(c,d)\,P_2$.

Similar to the finite-qubit case, translation-invariant Clifford unitaries
can be represented by symplectic transformations over the $R$-modules for Pauli operators.
That is, for a 2D lattice whose unit cell contains $q$ qubits,
translation-invariant Clifford unitaries are represented by
$V \in \Hom_R(R^{2q}, R^{2q})$ such that $V^\dagger \omega V = \omega$.
The most relevant examples in this work are those that do not mix $X$ and $Z$ operators:
\begin{equation}
    V = \begin{pmatrix}
        Q & 0 \\ 0 & (Q^\dagger)^{-1}
    \end{pmatrix},
    \label{eq: symp matrix for CNOT circuits}
\end{equation}
where $Q$ is an arbitrary invertible matrix over $R$, since we focus on CSS codes.
They can be realized by a constant-depth circuit with only CNOT gates
and ancillas.
We postpone the detailed discussion of circuit realization and ancillas to Section~\ref{sec: removing shifts}.

\paragraph{Codes on lattice.}
A translation-invariant topological stabilizer code on a 2D lattice
is defined by an abelian subgroup $\mathcal{S}$ (does not contain $-1$) of the Pauli group, 
called the \textit{stabilizer group}, satisfying the following conditions:
\begin{itemize}
    \item \textit{Translation invariance}: If $P \in \mathcal{S}$, then $T(a,b)\,P \in \mathcal{S}$ for all $(a,b)\in \mathbb{Z}^2$.
    Therefore, we can take some stabilizer generators such that all stabilizer generators can be obtained by translating one of them.
    Let the number of such generators be $m$.
    \item \textit{Commutativity}: $PQ = QP$ for all $P,Q\in\mathcal{S}$.
    \item \textit{Topological order}: For a Pauli operator $P$, 
    if $\forall Q \in \mathcal{S}$, $[P,Q]= 0$,
    then $P \in \mathcal{S}$.
    That is, there are no nontrivial local logical operators \cite{bravyiTopologicalQuantumOrder2010}.
\end{itemize}
Codes of this kind can be written in the language of modules over $R$ as follows.
Take $R^{2q}$ to represent the Pauli group as discussed above,
and take $R^m$ to represent the $m$ stabilizer generators.
Then we can assign a map $\sigma: R^m\to R^{2q}$ that encodes the 
information of these $m$ generators.
Therefore, the stabilizer group is now represented by the submodule $\im \sigma$.
With translation invariance holding automatically, the remaining two conditions impose the following requirements on $\sigma$:
\begin{itemize}
    \item \textit{Commutativity}: The submodule is isotropic: $\sigma^\dagger \omega\, \sigma = 0$.
    \item \textit{Topological order}: The submodule is coisotropic, and hence Lagrangian: $\ker(\sigma^\dagger \omega) = \operatorname{im} \sigma$.\footnote{
    Note that every element of $R$ is by definition a Laurent polynomial with finitely many terms,
    so in this language we always deal with finitely supported Pauli operators.
    This is why the topological order condition takes such a simple form.
    }
\end{itemize}
These two conditions are equivalent to the following chain complex 
being an exact sequence,
\begin{equation}
    R^m \xrightarrow{\;\sigma\;} R^{2q} \xrightarrow{\;\epsilon\, =\, \sigma^{\dagger}\omega\;} R^m,
    \label{eq: noncss sequence}
\end{equation}
as first formulated in Ref.~\cite{haahCommutingPauliHamiltonians2013}.
The map $\epsilon$ is called the excitation map and
describes how Pauli operators flip the stabilizer generators.
Therefore, $\im \epsilon$ consists of excited states that can be
generated from the ground state by local Pauli operators,
and $\coker \epsilon\coloneqq R^m/\im \epsilon$ consists of
classes of excited states;
each class contains states that can be transformed to each other
by local Pauli operators -- the anyons.
Fusion of anyons corresponds to addition in the cokernel.

The central object in this work is CSS codes,
for which the stabilizer group admits a generating set in which each generator contains exclusively $X$ or exclusively $Z$ operators.
For such codes, the maps $\sigma$ and $\epsilon$ adopt a block-diagonal form:
\begin{equation}
    \sigma = \begin{pmatrix}
        0_{q \times s} & H_Z^\dagger  \\  -H_X^\dagger &0_{q\times r} 
    \end{pmatrix},\quad 
    \epsilon = \begin{pmatrix}
        H_X & 0_{s\times q}   \\  0_{r\times q} & H_Z
    \end{pmatrix},
\end{equation}
with $r + s = m$.
The sequence~(\ref{eq: noncss sequence}) accordingly decomposes as a direct sum of two chain complexes,
\begin{equation}
    R^r\xrightarrow{H_Z^\dagger} R^q \xrightarrow{H_X} R^s,\quad 
    R^s\xrightarrow{-H_X^\dagger} R^q \xrightarrow{H_Z} R^r.
\end{equation}
Note that the minus sign in $-H_X^\dagger$ is inherited from
the way we map the commutation relation to the symplectic form.
Considering the structure of quantum codes,
since $\im H_X^\dagger = \im (-H_X^\dagger)$,
we can safely neglect the minus sign.
In what follows, we will call the second chain complex without the minus sign the ``dual'' of the first one.

Note that our notation is now parallel to
the usual chain complex notation for CSS codes,
with the following subtlety in interpreting the check matrix.
Over fields, one has two ways
to interpret a check matrix $H_X: R^q \to R^s$:
each row keeps the qubits involved in a check,
and each column keeps the checks flipped by a qubit.
However, here these two sets of information are related by
the spatial involution, so it is inevitable to break one of them.
We choose to keep the column interpretation,
in accordance with the interpretation of a check matrix as a map from bits to checks.
To read out the qubits involved in a check,
one should now look at columns of $H_X^\dagger$.

We further require that the stabilizer generators, as encoded in the specific form of $\sigma$, are independent,
i.e., that $H_X^\dagger$ and $H_Z^\dagger$ are injective.
It was shown in Ref.~\cite{haahCommutingPauliHamiltonians2013} that even if one starts with a non-injective $\sigma: R^{m'}\rightarrow R^{2q}$, 
another sequence describing the same code can be found whose $\sigma: R^m\rightarrow R^{2q}$ is injective. 
Here $m$ can be different from $m'$. 
We sidestep this subtlety and take injectivity as an assumption from here on.

To summarize, the object of interest in this work is a 2D translation-invariant topological CSS code defined by
the chain complex
\begin{equation}
    R^r\xrightarrow{H_Z^\dagger} R^q \xrightarrow{H_X} R^s,
    \label{eq: code chain}
\end{equation}
and its dual,
both being exact with injective $H_Z^\dagger$ and $H_X^\dagger$.
The following properties are proven in Ref.~\cite{haahCommutingPauliHamiltonians2013, haahAlgebraicMethodsQuantum2017} and we will take them as given:
\begin{itemize}
    \item $q = s + r$ --
    The ``density'' of independent checks equals the ``density'' of qubits.
    \item $\dim \coker H_X = \dim \coker H_Z = 0$, 
    i.e., $\dim_{\mathbb{F}_2} \coker H_X, \dim_{\mathbb{F}_2}\coker H_Z < \infty$ -- 
    There are finitely many anyons.
    \item $\dim_{\mathbb{F}_2}\coker H_X = \dim_{\mathbb{F}_2}\coker H_Z$ -- 
    The numbers of $e$- and $m$-type anyons are equal.
\end{itemize}

\begin{figure}
    \centering
    \includegraphics[width=.35\textwidth]{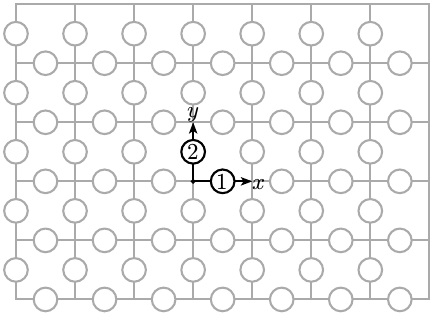}
    \hspace{2em}
    \includegraphics[width=.35\textwidth]{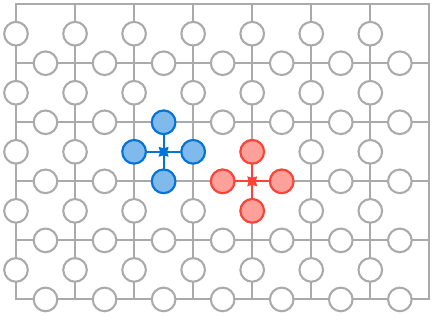}
    \caption{The toric code, with $q = 2$, $r = s = 1$.
    As shown in the left panel, there are two qubits per unit cell: the one on the horizontal edge (blue) is denoted by $(1,0)^\top$, and the one on the vertical edge (red) is denoted by $(0,1)^\top$.
    The only $X$-type stabilizer generator is the star operator $(\overline{x}-1,\overline{y}-1)^\top$ (the yellow one),
    and the only $Z$-type stabilizer generator is the plaquette operator $(1-y,x-1)^{\top}$ (the green one).
    }
    \label{fig:tc-demo}
\end{figure}

Finally, we present the example of fundamental importance in this work: the toric code.
See Fig.~\ref{fig:tc-demo}.
The check matrices are
\begin{equation}
    H_X = (x-1, y-1),\quad H_Z = (1-\overline{y},\overline{x}-1).
    \label{eq: toric code check matrix}
\end{equation}
For example, the $Z$ operator on qubit $(0,1)^\top$ violates the $X$-type stabilizer generators $y-1$, meaning the generators on the origin and on $(0,1)$.

In this work, all results are explicitly written for qubit systems.
Nevertheless, they can be naturally generalized to
qudit systems with prime dimension $p$.
For non-prime dimensions, new subtleties arise as explained in Ref.~\cite{SongNew}.

\section{Comments on algebraic methods in this work}
\label{app:algebraic-methods}

In this section, we review two trends of algebraic methods 
for studying quantum codes
to position our work within the landscape 
of existing results and future directions of research.

The starting point of the algebraic methods reviewed here
is the formulation of quantum CSS codes as chain complexes \cite{freedman2002z2, kitaevFaulttolerantQuantumComputation2003, bravyiQuantumCodesLattice1998}.
The key observation is that commutativity of stabilizers imposes
the condition $H_X H_Z^\top = 0$ on check matrices,
allowing one to write a quantum CSS code with $n$ physical qubits,
$c_Z$ $Z$-checks and $c_X$ $X$-checks as a chain complex
\begin{equation}
    \mathbb{F}_2^{c_Z}\xrightarrow{H_Z^\top} \mathbb{F}_2^n \xrightarrow{H_X} \mathbb{F}_2^{c_X}.
    \label{eq:css-field-chain}
\end{equation}
Logical Pauli $Z$ operators are represented by the homology group $\ker H_X/\im H_Z^\top$,
and logical Pauli $X$ operators are represented by the cohomology group.

Given this formulation, one can therefore study the morphisms between
chain complexes to study different aspects of quantum codes.
This is the first trend that forms the methodological background of our work.
Some developments in this direction are listed as follows.
\begin{itemize}
    \item Chain isomorphisms imply unitaries that map one code to another.
    Specifically, if the unitary satisfies some locality conditions,
    it implements fault-tolerant unitary code switching.
    \item Automorphisms imply intra-block logical Clifford gates,
    such as fold-transversal gates \cite{breuckmannFoldTransversalCliffordGates2024, eberhardtLogicalOperatorsFoldtransversal2024}.
    \item Chain homomorphisms imply inter-block logical Clifford gates,
    such as the homomorphic CNOT gates \cite{huangHomomorphicLogicalMeasurements2022, xuFastParallelizableLogical2024}.
    \item Logical measurements based on lattice surgery techniques \cite{horsmanSurfaceCodeQuantum2012, cohenLowoverheadFaulttolerantQuantum2022}
    can be formulated as mapping cones \cite{ideFaulttolerantLogicalMeasurements2024}.
\end{itemize}
In this work, we focus on unitary equivalence between quantum codes,
for which chain isomorphisms are the most relevant.

To impose symmetries on codes,
the action of symmetry groups should be considered in the algebraic formulation.
This is the second trend that forms the methodological background of our work.
Let the symmetry group be $G$. 
Then the vector spaces in Eq.~\eqref{eq:css-field-chain}
should now be viewed as $\mathbb{F}_2G$-modules,
and the linear maps should now be $\mathbb{F}_2G$-homomorphisms.
If we take $G = \mathbb{Z}^2$, the translation symmetry of a 2D lattice,
the group algebra $\mathbb{F}_2G$ becomes the Laurent polynomial ring $R = \mathbb{F}_2[x^{\pm 1}, y^{\pm 1}]$.
This is exactly the formulation we reviewed in Section \ref{sec: fundamental}.
For general symmetries, this formulation leads to constructions
of new codes including the balanced product codes \cite{breuckmannBalancedProductQuantum2021}
and lifted product codes \cite{panteleevQuantumLDPCCodes2022},
which lead to the construction of good quantum low-density parity-check codes \cite{panteleevAsymptoticallyGoodQuantum2022}.

In summary, our work lies in a special intersection of these two trends:
we study isomorphisms between chain complexes over $R$-modules.
Techniques from homological algebra and 
from commutative algebra are therefore necessary in our work.
More intersections of these two trends,
that is, to study constructions of gates, measurements or decoders
with the knowledge of symmetries of the codes,
can be left for future exploration.
We will comment on some such topics as applications of our result
in Section~\ref{sec: applications}.

\section{Coarse-graining and superlattice}

In this section,
we discuss the anyon-permuting properties 
of the translation symmetry and coarse-graining.
In Subsection \ref{subsec: review wsb}, known results
relating to the anyon permutations are reviewed,
with a description of how to compute the superlattice.
The physics and algebra of coarse-graining is also discussed,
serving as the starting point of our main results.
In Subsection \ref{subsec: bound}, we derive an upper bound of anyon periods, 
partially answering the problem raised in Ref.~\cite{haahClassificationTranslationInvariant2021}.

Throughout this section, all results are stated explicitly for CSS codes, since our main decoupling algorithm is designed for that setting.
We will always discuss the exact sequence \eqref{eq: code chain},
which is related to $Z$ operators and $e$ anyons.
Properties for $X$ operators and $m$ anyons follow from
the duality $H_X \leftrightarrow H_Z$.
These results also generalize to arbitrary Pauli stabilizer codes 
by directly considering the non-CSS chain complex \eqref{eq: noncss sequence}.

\subsection{Coarse-graining}\label{subsec: review wsb}

The notion of weak symmetry breaking,
or anyon-permutation symmetry,
describes the phenomenon that symmetry operations that
leave the Hamiltonian and ground state invariant 
can permute anyons \cite{kitaevAnyonsExactlySolved2006, barkeshliSymmetryFractionalizationDefects2019}.
It is observed in various systems, 
including the Wen plaquette model
\cite{wenQuantumOrdersSymmetric2002, wenQuantumOrdersExact2003}
(equivalently, rotated surface codes
\cite{bombinOptimalResourcesTopological2007}),
color codes \cite{bombinStatisticalMechanicalModels2008},
and non-stabilizer models like the $A$-phase (gapped phase)
of Kitaev's honeycomb lattice model \cite{kitaevAnyonsExactlySolved2006}.

\begin{figure}
    \centering
    \includegraphics[width=0.3\textwidth]{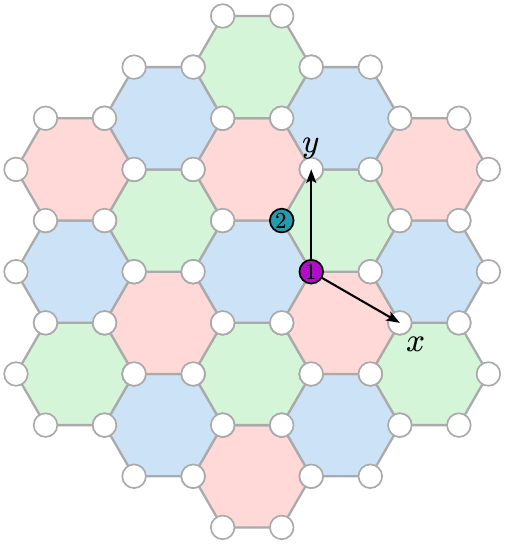}
    \hspace{2em}
    \includegraphics[width=0.3\textwidth]{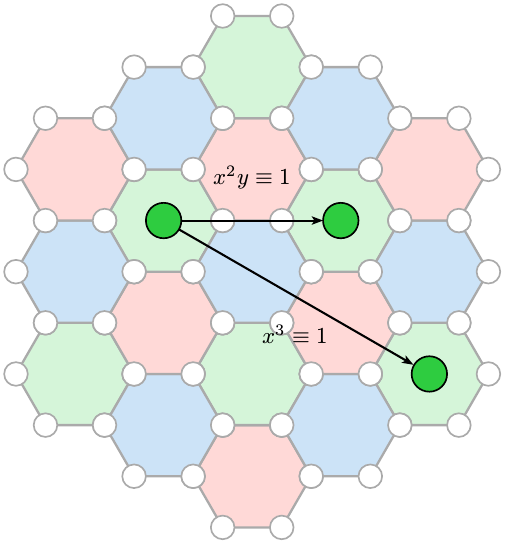}
    \caption{The 6.6.6 color code, with $q = 2$, $r = s = 1$.
    As shown in the first panel, there are two qubits per unit cell: the purple one is denoted by $(1,0)^\top$, while the green one is denoted by $(0,1)^\top$.
    This code is self-dual: both $X$- and $Z$-type stabilizer generators are the weight-6 plaquette operators.
    The anyons are classified by their ``colors'', 
    and the anyon-preserving translation symmetry is generated by $x^3$ and $x^2y$, as shown in the second panel.
    }
    \label{fig:sl-666cc}
\end{figure}

In this paper, we are mostly interested in
the weak breaking of the translation symmetry.
It is straightforward to see this phenomenon through
the example of $6.6.6$ color codes.
In a $6.6.6$ color code, qubits are put on vertices of a hexagonal lattice,
and an $X$-type stabilizer generator and a $Z$-type stabilizer generator
are assigned to each hexagon.
See Fig.~\ref{fig:sl-666cc} for an illustration.
Colors in the color code label the anyons:
the $e$-type anyons in a color code are described by the space $\mathbb{F}_2^2$,
and the three non-trivial anyons are ``colored'' in red, green and blue, respectively.
From Fig.~\ref{fig:sl-666cc} one can see that translation by a lattice vector
changes the color of an anyon, and thus permutes the anyons.
It also illustrates that by switching to a coarser lattice
generated by $x^2y$ and $x^3$ we can make this permutation trivial.
For a detailed discussion on this example (together with its decoupling unitary),
see Section~\ref{sec: examples}.

In the general case, the anyon-permutation properties of translations
are manifested by the action 
of the ring $R$ on the anyon module $\coker H_X$.
Since $\coker H_X$ is a $t$-dimensional vector space,
invertible variables $x,y$ act as invertible matrices $X,Y\in \mathrm{GL}(t,\mathbb{F}_2)$.
This defines a representation
$\omega_1:\mathbb{Z}^2\to \mathrm{GL}(t, \mathbb{F}_2)$
of the translation group on the anyon space.
Therefore, translations that leave the anyons invariant
are those in the subgroup $\ker \omega_1 \leq \mathbb{Z}^2$.
Since $\mathrm{GL}(t, \mathbb{F}_2)$ is finite,
$\ker \omega_1$ is isomorphic to $\mathbb{Z}^2$,
which is usually referred to as the \textit{superlattice}.
The unit cell of the superlattice is called a \textit{supercell},
whose size with respect to the old unit cell equals
the index $[\mathbb{Z}^2\colon \ker \omega_1]$ of the subgroup.
Algorithmically, the generators of the superlattice
can be found by computing a monomial basis for each of $\coker H_X$ and $\coker H_Z$ using Gr\"obner basis techniques \cite{coxUsingAlgebraicGeometry2005},
computing the translation representation on their direct sum, and traversing the lattice vectors
to find an anyon-non-permuting pair that spans the minimal area.

As is discussed in the main text,
decoupling is possible only when we rewrite 
the chain complex of the code in the superlattice
so that translations do not permute anyons.
Now we discuss the consequence of such rewriting.
These are the mathematical conditions in our main theorem.

First, there is a change in the parameters $r, q, s$,
each increased by a factor of $\Lambda = [\mathbb{Z}^2\colon \ker \omega_1]$.
We still write the new chain complex in the form of Eq.~\eqref{eq: code chain}.
For example, for the color code discussed above,
the parameters are now $q = 6$ and $r = s = 3$.
The complexity stated in the main theorem depends on this $q$,
which is proportional to the size of the supercell
rather than the $q$ in the original chain complex.

Second, we have the annihilator condition \cite{haahCommutingPauliHamiltonians2013, haahAlgebraicMethodsQuantum2017}
\begin{equation}
    \ann \coker H_X = \ann \coker H_Z = (x-1,y-1).
    \label{eq: ann coker}
\end{equation}
By definition, $(x-1,y-1)\subseteq \ann \coker H_X$ means
for any $e \in R^s$, there exists $u, v \in R^q$ such that 
$H_X u = (x-1) e$ and $H_X v = (y-1) e$.
This is exactly the algebraic way of saying
any local excitation $e$ can be transported
by one lattice vector by some ``hopping operator'' $u$ or $v$,
which holds in the superlattice by construction.
Equality holds since $\ann \coker H_X \neq R$ (there exist non-trivial anyons)
and $(x-1,y-1)$ is maximal.

\begin{remark} In Ref.~\cite{haahCommutingPauliHamiltonians2013},
Lemma 7.3, Haah gave an algebraic proof of the existence of the
anyon-preserving superlattice.
In Haah's proof, the annihilator ideal $I = \ann(\coker \epsilon)$ plays the central role.
Given that $\dim_{\mathbb{F}_2}(R/I)$ is finite (equivalently, $R/I$ has Krull dimension zero), he shows that there exists an integer $L$ such that $(x^L - 1, y^L - 1) \subseteq I$.
He accomplishes this by considering each maximal ideal $\mathfrak{m} \supseteq I$ and finding a local period $L_{\mathfrak{m}}$ such that $(x^{L_{\mathfrak{m}}} - 1, y^{L_{\mathfrak{m}}} - 1) \subseteq \mathfrak{m}$, then combining these local periods into a single global period $L$.

This algebraic proof is related to the physical picture
of anyon permutation in the following sense:
each maximal ideal corresponds to a subspace of anyons
that is invariant under the action
of translation matrices $X, Y \in \mathrm{GL}(t,\mathbb{F}_2)$.
To see this, note that since $\dim_{\mathbb{F}_2}(\coker \epsilon)$ is finite, the quotient $R/I$ is an Artin ring.
By the structure theorem for Artin rings \cite[Theorem~8.7]{atiyah1994introduction} (a multivariable analogue of the Chinese Remainder Theorem), we decompose $R/I \cong \bigoplus_i (R/I)_{\mathfrak{m}_i}$, where the $\mathfrak{m}_i$ are the distinct maximal ideals of $R/I$.
This decomposition is inherited by the $(R/I)$-module $\coker \epsilon$:
\begin{equation}
\coker \epsilon \cong\bigoplus_i (\coker \epsilon)_{\mathfrak{m}_i}.
\end{equation}
The direct-sum structure ensures invariance under the action of both $x$ and $y$, so this decomposition is precisely the simultaneous block-diagonalization of $X$ and $Y$.
The $i$-th block isolates a generalized joint eigenspace of the translation operators.
Restricted to this subspace, the annihilating polynomials of the translation operators are exactly the elements of the ideal $\mathfrak{a}_i$ satisfying $(R/I)_{\mathfrak{m}_i} \cong R/\mathfrak{a}_i$.

\end{remark}

\subsection{Bounding the supercell size by interaction range}\label{subsec: bound}

In Ref.~\cite{haahClassificationTranslationInvariant2021}, 
Haah raised an open question to bound the length
of a translation vector that leaves anyons invariant,
or equivalently, to bound the number of qubits in the supercell.
More precisely, given a family of 2D translation-invariant Hamiltonians describing topological stabilizer codes with fixed unit cell size $q$ and interaction range $l$ (in accordance with physical intuition, defined below),
what is the largest possible supercell?
Haah claimed that this upper bound should be no less than $\exp \Omega(l)$ \cite{haahCommutingPauliHamiltonians2013}.
In this section, we give an upper bound of $\exp(\mathcal{O}(q^3 l^2))$,
which, as we will argue below, may not be a tight bound.

We arrive at this bound in two steps.
First, we bound the superlattice index by the number of anyons,
which is a direct result of our discussion of matrix representations
of translation operators.
Second, we bound the number of anyons
in terms of Hamiltonian parameters, 
motivated by the techniques introduced in Ref.~\cite{chenAnyonTheoryTopological2025}.
These two individual bounds need not be saturated simultaneously,
leading to a potential gap between our bound and the tight bound.

To bound the superlattice index by the number of anyons,
we need the following result from linear algebra,
a direct corollary of the Cayley--Hamilton theorem.

\begin{lemma}\label{lemma: order bound linear operator}
    Let $A\in \mathrm{GL}(t,\mathbb{F}_2)$.
    Then the order of $A$, i.e., the smallest positive integer $n$ such that $A^n = I$,
    is at most $2^t-1$.
\end{lemma}

\begin{proof}
    Let $p(\lambda)$ be the characteristic polynomial of $A$.
    By the Cayley--Hamilton theorem, $p(A) = 0$,
    so $A^t$ is a linear combination of $I, A, \dots, A^{t-1}$.
    Hence the subalgebra of $\mathrm{Mat}(t,\mathbb{F}_2)$ generated by $A$ has dimension at most $t$.
    The set $\{A^n : n \in \mathbb{Z}\}$ is contained in this subalgebra
    and cannot contain the zero matrix,
    so it has at most $2^t - 1$ elements.
\end{proof}

The next step is to bound the number of anyons by Hamiltonian parameters.
Note that here all parameters belong to the original Hamiltonian,
which are the input data we use to compute the superlattice.
For example, the parameter $q$ is the number
of qubits per unit cell \textit{before} coarse-graining.
The interaction range $l$ can be defined as follows.
Intuitively, we say a Hamiltonian has interaction range $l$
if each of its interaction terms is supported by qubits
that lie in an $l\times l$ square.
Algebraically, for the $X$-part of the CSS code,
a term in the Hamiltonian is a column in $H_X^\dagger$.
To determine the ``range'' of the support of such a term,
we plot the exponent vectors ($(m,n)$ for $x^m y^n$)
of all monomials appearing in this column,
and find a minimal square to cover them.
In other words, 
$l-1$ is the maximum of the differences of degrees in $x$ and in $y$.
(For example, $H_X^\dagger=(x-1, y-1)^\dagger$ yields $l = 2$, with the maximum difference being 1.)
Since this characterization yields the same result 
under taking antipodes $x\mapsto \overline{x}$, $y\mapsto \overline{y}$,
we can access the interaction range from the rows of $H_X$ as well as columns of $H_X^\dagger$.
With the characterization of Hamiltonian parameters,
the following lemma accomplishes the second step.

\begin{lemma}\label{lemma: anyon number}
Let $R = \mathbb{F}_2[x^{\pm 1}, y^{\pm 1}]$ be a Laurent polynomial ring, and let $H\colon R^m \to R^n$ be an $R$-linear map whose cokernel has Krull dimension zero.
Suppose that for each row of the matrix representation of $H$, the exponent vectors in $\mathbb{Z}^2$ of all monomials appearing in that row lie within a square of side length $l$.
Then the $\mathbb{F}_2$-dimension of the cokernel satisfies
\begin{equation}
\dim_{\mathbb{F}_2} (\coker H) \le 2 n^3 l^2.
\end{equation}
\end{lemma}
\begin{proof}
Let $I = \operatorname{Fitt}_0(\coker H)$ be the zeroth Fitting ideal of $\coker H$.
Since $I \subset \ann(\coker H)$, there is a natural surjection $(R/I)^n \twoheadrightarrow \coker H$, giving
\begin{equation}
\dim_{\mathbb{F}_2}(\coker H) \le n \cdot \dim_{\mathbb{F}_2}(R/I).
\end{equation}
It remains to bound $\dim_{\mathbb{F}_2}(R/I)$.
Each $n \times n$ minor determinant is a sum of products containing exactly one entry per row, so its Newton polygon lies in the Minkowski sum of $n$ copies of a square of side length $l$ --- that is, a square $S$ of side length $nl$.
Passing to the algebraic closure $K = \bar{\mathbb{F}_2}$, the extended ideal $I_K = I \otimes_{\mathbb{F}_2} K$ remains zero-dimensional, and a generic pair of elements $f, g \in I_K$ forms a regular sequence whose Newton polygons are contained in $S$.
By Bernstein's theorem~\cite{bernshteinNumberRootsSystem1975},
\begin{equation}
\dim_K \bigl(K[x^{\pm 1}, y^{\pm 1}]/(f,g)\bigr) \le \mathcal{M}(S, S) = 2\,\operatorname{Area}(S) = 2(nl)^2.
\end{equation}
Since $\dim_{\mathbb{F}_2}(R/I) \le \dim_K \bigl(K[x^{\pm 1}, y^{\pm 1}]/(f,g)\bigr)$, substituting into the first inequality yields the claimed bound.
\end{proof}

Applying the lemmas to CSS codes, we have the following theorem.

\begin{theorem}\label{thm:index-bound}
    Consider a translation-invariant topological CSS code with $q$ qubits, $r$ types of independent $Z$-type stabilizer generators, and $s$ types of independent $X$-type stabilizer generators per unit cell (with $q = r + s$).
    Suppose each $X$-type stabilizer generator is supported within an $(l_X+1)\times(l_X+1)$ square on the lattice, and each $Z$-type stabilizer generator within an $(l_Z+1)\times(l_Z+1)$ square.
    Then the anyon space has $\mathbb{F}_2$-dimension
    \begin{equation}
       a \leq 4\min\left\{r^3 l_Z^2, s^3 l_X^2\right\},
    \end{equation}
    If $q'$ denotes the number of qubits in a supercell of the anyon-preserving superlattice, then
    \begin{equation}
        q' \leq q(2^a-1)^2.
    \end{equation}
\end{theorem}

\begin{proof}
    By the discussion before Lemma~\ref{lemma: anyon number},
    we can apply Lemma~\ref{lemma: anyon number} to $H_X$ to obtain $\dim_{\mathbb{F}_2}(\coker H_X) \leq 2s^3 l_X^2$.
    Similarly we have $\dim_{\mathbb{F}_2}(\coker H_Z) \leq 2r^3 l_Z^2$.
    Since the $e$- and $m$-type sector counts coincide, each is bounded by $\min\{2s^3 l_X^2, 2r^3 l_Z^2\}$, so the total satisfies $a \leq 4\min\{r^3 l_Z^2, s^3 l_X^2\}$.
    By Lemma~\ref{lemma: order bound linear operator}, the size of a supercell relative to the original unit cell is bounded by the index of the superlattice
    $\{(m,n): X^m = I, Y^n = I\}$,
    which is the product of the orders of $X$ and $Y$ and hence is at most $(2^a-1)^2$.
    Multiplying this relative size by the original number $q$ of qubits per unit cell gives the stated bound on $q'$.
\end{proof}

\begin{remark}
Let $q=r+s$ be the number of qubits per unit cell before coarse-graining,
and set $l=\max\{l_X,l_Z\}$. Since
$\min\{r,s\}\le q/2$, Theorem~\ref{thm:index-bound} gives
\begin{equation}
    a \le 4\min\{r^3,s^3\}l^2
    \le \frac{q^3l^2}{2}.
\end{equation}
Consequently, the number of qubits in an anyon-preserving supercell satisfies
\begin{equation}
    q' \le q(2^a-1)^2 < q2^{2a}\le q2^{q^3l^2},
\end{equation}
which is the bound quoted in the main text.
\end{remark}

\section{The main results}

In this section, we establish the main theorem of this work.
We prove the existence of an explicit decoupling map
specifying the Clifford unitary
that transforms any translation-invariant topological CSS code
into stacks of toric codes and product states
while respecting the maximal residual translation symmetry,
and present the construction
of the polynomial-time algorithm that produces this map.
The map can be compiled to a constant-depth Clifford circuit 
via the Suslin stability theorem;
see Sec.~\ref{sec: removing shifts} for a detailed discussion.

We begin by presenting in full detail
our main theorem and algorithm,
with a discussion on the difficulties.
The remaining three subsections follow the strategy
introduced in the main text: 
stripping off cokernels,
proving the existence
and finishing the construction.

\subsection{Main theorem}\label{sec: stating problem}

Our main theorem is stated as follows.

\begin{theorem}[Symmetry-preserving unitary decoupling]\label{thm: main}
Given a code whose chain complex and its dual
\begin{equation}
    R^r\xrightarrow{H_Z^\dagger} R^q \xrightarrow{H_X} R^s,\quad  R^s\xrightarrow{H_X^\dagger } R^q \xrightarrow{H_Z } R^r
\end{equation}
have the following properties:
\begin{enumerate}
    \item both complexes are exact;
    \item $H_Z^\dagger, H_X^\dagger$ are injective;
    \item $\dim_{\mathbb{F}_2} \coker H_X =\dim_{\mathbb{F}_2}\coker H_Z= t<\infty$,
and  $\ann\coker H_Z = \ann \coker H_X = (x-1,y-1)$,
\end{enumerate}
then there exists a chain isomorphism between the first chain complex and
a $(t,r,s)$-standard chain complex representing $t$ copies of toric codes, together with
$p_Z = r - t$ copies of $Z$-basis product states,
and $p_X = s - t$ copies of $X$-basis product states
with check matrices
\begin{equation}
    \begin{aligned}
        \tilde H_X &= \begin{pmatrix}
            I_{p_X} & 0_{p_X\times p_Z} & 0_{p_X\times t} & 0_{p_X\times t} \\
            0_{t\times p_X} & 0_{t\times p_Z} & (x-1) I_t & (y-1)I_t
        \end{pmatrix},\\
        \tilde H_Z^\dagger &= \begin{pmatrix}
            0_{p_X\times p_Z} & 0_{p_X\times t} \\ 
            I_{p_Z} & 0_{p_Z\times t} \\ 
            0_{t\times p_Z} & (1-y)I_{t} \\ 
            0_{t\times p_Z} & (x-1)I_{t} \\ 
        \end{pmatrix}.
    \end{aligned}
    \label{eq: goal matrices}
\end{equation}
That is, there exist $R$-isomorphisms $(\psi_2, \psi_1, \psi_0)$ such that the following diagram commutes:
\begin{equation}
\begin{tikzcd}
	{\textnormal{input:}} & {R^{r}} & {R^{q}} & {R^s} \\
	{\textnormal{standard:}} & {R^{r}} & {R^{q}} & {R^s}
	\arrow["{{{H_Z^\dagger}}}", from=1-2, to=1-3]
	\arrow["{{{H_X}}}", from=1-3, to=1-4]
	\arrow["\psi_2", from=1-2, to=2-2]
	\arrow["{{{\tilde{H}_Z^\dagger}}}"', from=2-2, to=2-3]
	\arrow["{\psi_1}", from=1-3, to=2-3]
	\arrow["{{{\tilde{H}_X}}}"', from=2-3, to=2-4]
	\arrow["{{{\psi_0}}}", from=1-4, to=2-4]
\end{tikzcd}
\label{eq: chain map}
\end{equation}
The following diagram then automatically commutes for the dual chain:
\begin{equation}
\begin{tikzcd}
	{R^{s}} & {R^{q}} & {R^r} \\
	{R^{s}} & {R^{q}} & {R^r}
	\arrow["{{{{H_X^\dagger}}}}", from=1-1, to=1-2]
	\arrow["{(\psi^\dagger_0)^{-1}}", from=1-1, to=2-1]
	\arrow["{{{{H_Z}}}}", from=1-2, to=1-3]
	\arrow["{{(\psi^\dagger_1)^{-1}}}", from=1-2, to=2-2]
	\arrow["{{{{(\psi^\dagger_2)^{-1}}}}}", from=1-3, to=2-3]
	\arrow["{{{{\tilde{H}_X^\dagger}}}}"', from=2-1, to=2-2]
	\arrow["{{{{\tilde{H}_Z}}}}"', from=2-2, to=2-3]
\end{tikzcd}
\end{equation}
Furthermore, Algorithm~\ref{alg: constructing isomorphism}
computes these isomorphisms in time polynomial in $q$ and
the degrees of the input check matrices $H_Z, H_X$.
\end{theorem}
\begin{algorithm}[htb]
\SetAlgoLined
\KwIn{check matrices $H_Z, H_X$}
\KwOut{invertible maps $(\psi^{-1}_2, \psi^{-1}_1, \psi^{-1}_0)$}
Find
$\xi_2\in\mathrm{GL}(r,\mathbb F_2)$,
$\xi_1\in\mathrm{GL}(q,\mathbb F_2)$, and
$\xi_0\in\mathrm{GL}(s,\mathbb F_2)$
such that $H^\dagger_Z \xi_2 = \xi_1\tilde H_Z^\dagger \pmod J$ and $H_X \xi_1 = \xi_0 \tilde H_X \pmod J$ using standard linear algebra over $\mathbb{F}_2$\;
$H_X \gets \xi_0^{-1}H_X\xi_1$, $H_Z^\dagger\gets \xi_1^{-1}H_Z^\dagger \xi_2$\;
Solve $\tilde{H}_X = H_X \phi_1$ for $(\phi_1)_{q\times q}$, with the columns $[p_X + 1:p_X + p_Z]$ of $\phi_1$ taken from the first $p_Z$ columns of $H_Z^\dagger$\;
Solve $H^\dagger_Z \phi_2 = \phi_1 \tilde{H}_Z^\dagger$
for $(\phi_2)_{r\times r}$
\;
Replace the lower-right $t\times t$ corner $B$ of $\phi_2$ by $B\bmod J$ to obtain $\phi_2'$\;
Solve $\eta\tilde{H}_Z^\dagger = \phi_2'-\phi_2$
for $\eta_{r\times q}$
\;
$\phi_1' \gets \phi_1 + H_Z^\dagger \eta$\;
\Return $(\xi_2\phi_2', \xi_1 \phi_1', \xi_0)$\;
\caption{Solving for the decoupling unitary}
\label{alg: constructing isomorphism}
\end{algorithm}

Here we comment on the physical meaning of the isomorphisms $(\psi_2, \psi_1, \psi_0)$.
The related equations are 
\begin{equation}
    \tilde{H}_X \psi_1 = \psi_0 {H}_X,\quad \tilde{H}_Z^\dagger \psi_2 = \psi_1 H_Z^\dagger.
\end{equation}
Let $U$ be the Clifford decoupling unitary.
For any $u\in R^q$ representing a Pauli $Z$ string $P_Z$ on the input code,
$U P_Z U^\dagger$ is represented by $\psi_1u$.
As for the dual maps, for any $v\in R^q$ representing 
a Pauli $X$ string $P_X$ on the input code, 
$U P_X U^\dagger$ is represented by $(\psi_1^{-1})^\dagger v$.
The invertible maps $\psi_2$ and $\psi_0$ describe how stabilizer generators and excitations are mapped.
Once $\psi_1$ is obtained, the corresponding CNOT circuit with lattice shifts can be compiled by decomposing $\psi_1$ into elementary matrices using Suslin stability~\cite{suslinStructureSpecialLinear1977, parkAlgorithmicProofSuslins1994},
which we will discuss in more detail in Sec.~\ref{sec: removing shifts}.
We are not aware of a compact bound on the time complexity of the algorithm for Suslin stability \cite{parkAlgorithmicProofSuslins1994};
the polynomial-time claim in Theorem~\ref{thm: main} therefore only refers to constructing the maps $(\psi_2,\psi_1,\psi_0)$,
hence to the unitary for decoupling, but not to compiling the explicit CNOT circuit.

Before giving a complete proof and construction,
we also comment here on the technical difficulties underlying this problem.
At first glance, the decoupling problem appears to reduce to transforming matrices into a standard form,
solvable by Gaussian elimination or linear algebra over $R$.
However, naive approaches face intrinsic difficulties.

First, Gaussian-type elimination algorithms suffer from severe intermediate expression swell.
When clearing a column, one may multiply a row by a monomial taken from another row, roughly doubling the degree.
In the worst case, the maximal degree of the matrix entries can therefore grow exponentially.
In Haah's algorithmic proof~\cite{haahAlgebraicMethodsQuantum2017},
an inductive method is used to control the intermediate swell:
toric-code sectors are brought into standard form one at a time,
with only multiplication by polynomials of uniformly bounded degree allowed at each step.
This method requires iterative coarse-graining,
so the breaking of translation symmetry is not tracked~\cite{haahClassificationTranslationInvariant2021}.

A natural workaround is to solve for the transformation directly using
computer algebra techniques such as Gr\"obner bases for submodules~\cite{coxUsingAlgebraicGeometry2005}.
However, for linear systems with non-trivial kernels,
an invertible (isomorphic) solution is not guaranteed by default.
In ordinary linear algebra over a field, one would fix this by decomposing the ambient space
as a direct sum of the kernel and a complementary subspace,
then finding invertible solutions for both spaces
and taking a direct sum.
Over $R$, however, this strategy also fails:
the kernel of a generic $R$-homomorphism may not be a direct summand of
some free modules.

We can see this from the following minimal example.
For $H_X = (x-1, y-1)$, the standard toric code (see Eq.~\eqref{eq: toric code check matrix}), we have
\begin{equation}
    \begin{pmatrix}
       x-1 & y-1 
    \end{pmatrix} = 
    \begin{pmatrix}
       x-1 & y-1 
    \end{pmatrix}
    \begin{pmatrix}
        1 & x - xy \\ 
        0 & 1 - x + x^2
    \end{pmatrix}.
\end{equation}
Even though the identity matrix is an invertible solution,
a generic solver can still output a result of this type
for more complicated cases.
The reason is that $\ker H_X = \im H_Z \subset R^2$.
One can always add an element from the kernel to a column of the solution,
and there is no submodule $M$ of $R^2$ such that $R^2 = \ker H_X \oplus M$.

A celebrated approach to ``decoupling'' the kernel of a map between free modules is the Quillen--Suslin theorem~\cite{quillenProjectiveModulesPolynomial1976, suslinProjectiveModulesPolynomial1976}.
Although it is not applicable to the full problem because $\im H_X$ is generally not projective, Haah suggested using it for the product-state part of the decoupling problem \cite{haahAlgebraicMethodsQuantum2017}.
Furthermore, we are not aware of any polynomial time-complexity result for such algorithms,
and if the coefficient field is infinite,
such algorithms can lead to complexity exponential in the matrix size $q$ \cite{canigliaAlgorithmicAspectsSuslins1993}.

We now present our proof in three steps corresponding to the sketch in the main text.
For technical reasons, the natural outputs of our construction are the inverses of the isomorphisms $(\psi_2,\psi_1,\psi_0)$.
Inverting matrices over polynomial or Laurent rings with a fixed number of variables
is standard in computer algebra and can be implemented with time complexity polynomial in the matrix size and the original degree;
the degree of the inverted matrix is also polynomial in the matrix size and the original degree \cite{coxUsingAlgebraicGeometry2005, liang2022degree}.
Therefore, we omit the matrix-inversion step in our proof and in Algorithm~\ref{alg: constructing isomorphism}.

\subsection{Step 1: Stripping off cokernels}

First, we strip off the cokernels to write 
the sequences to be transformed into short exact sequences.
Intuitively, it is equivalent to putting the code on a torus 
by imposing periodic boundary conditions on a supercell
and extracting the $2t$ logical qubits to $2t$ physical ones.
Putting this decoupling unitary back on the infinite plane,
it pins down the positions of qubits 
that will become $Z$- and $X$-basis product states after the full decoupling.
However, we will give a more elementary proof here,
and delay the full discussion of periodic boundary conditions to Section~\ref{subsec: pbc}.

\begin{lemma}\label{lem: clear in F2}
Given an exact sequence of a code,
\begin{equation}
    R^r\xrightarrow{H_Z^\dagger} R^q \xrightarrow{H_X} R^s,
\end{equation}
with injective $H_Z^\dagger, H_X^\dagger$,
$\dim_{\mathbb{F}_2} \coker H_X = t$,
and $\ann\coker H_Z = \ann \coker H_X = (x-1,y-1)$,
    there exist invertible $\mathbb{F}_2$ matrices $(\xi_2)_{r \times r}$, $(\xi_0)_{s\times s}$, and $(\xi_1)_{q\times q}$ such that
    \begin{equation}
        H^\dagger_Z \xi_2 = \xi_1\tilde H_Z^\dagger \pmod J,\quad 
        H_X \xi_1 = \xi_0 \tilde H_X \pmod J,
        \label{eq: clear H mod J}
    \end{equation}
    and such that the sequence 
    \begin{equation}
        0\rightarrow R^r \xrightarrow{\xi_1^{-1}H^\dagger_Z \xi_2} R^q \xrightarrow{\xi_0^{-1}H_X\xi_1} R^{p_X}\oplus JR^t \rightarrow0
        \label{eq: ses std mod J}
    \end{equation}
    is well-defined and exact.
\end{lemma}

\begin{proof}
    The lemma is equivalent to the following statement:
    tensoring the rows of~(\ref{eq: chain map}) in Theorem~\ref{thm: main} with $R/J = \mathbb{F}_2$,
    i.e., reducing modulo $J$,
    there exist invertible $(\xi_2, \xi_1, \xi_0)$ as stated such that the following diagram commutes:
    \begin{equation}
        \begin{tikzcd}
        	0 & {\mathbb{F}_2^{r}} & {\mathbb{F}_2^{q}} & {\mathbb{F}_2^s} \\
        	0 & {\mathbb{F}_2^{r}} & {\mathbb{F}_2^{q}} & {\mathbb{F}_2^s}
        	\arrow[from=1-1, to=1-2]
        	\arrow["{(H_Z^\dagger)_\ast}", from=1-2, to=1-3]
        	\arrow["{(H_X)_\ast}", from=1-3, to=1-4]
        	\arrow[from=2-1, to=2-2]
        	\arrow["\xi_2", from=2-2, to=1-2]
        	\arrow["{(\tilde{H}_Z^\dagger)_\ast}"', from=2-2, to=2-3]
        	\arrow["\xi_1", from=2-3, to=1-3]
        	\arrow["{(\tilde{H}_X)_\ast}"', from=2-3, to=2-4]
        	\arrow["{\xi_0}", from=2-4, to=1-4]
        \end{tikzcd}
        \label{eq: clear sequence mod J}
    \end{equation}
    This is because any isomorphism between $\mathbb{F}_2$-vector spaces
    can be lifted to an $R$-isomorphism between free $R$-modules,
    with the same matrix representation (we will comment on this point further after the proof).
    To show the existence of these maps,
    it suffices to show that the two rows have the same dimension
    for corresponding homology vector spaces.

    For the second row, we can read from matrices that (dim refers to dimension of vector spaces in this proof)
    $\dim \coker (\tilde H_X)_\ast = t$, $\dim (\ker (\tilde H_X)_\ast / \im (\tilde H_Z^\dagger)_\ast) = 2t$, and $\dim \ker (\tilde H_Z^\dagger)_\ast = t$.
    For the first row, $\ann \coker H_X = J$ implies $JR^s\subseteq \im H_X\subseteq R^s$.
    Therefore $\coker (H_X)_\ast = (R^s/JR^s) / (\im H_X/JR^s) \cong R^s/\im H_X = \coker H_X$,
    leading to $\dim \coker (H_X)_\ast = t$.
    Similarly we have for linear map $(H_Z^\dagger)_\ast$,
    $\dim \ker (H_Z^\dagger)_\ast = \dim \coker (H_Z)_\ast = t$.
    Finally, $\dim (\ker (H_X)_\ast / \im (H_Z^\dagger)_\ast) = \dim \ker (H_X)_\ast - \dim \im (H_Z^\dagger)_\ast = (q - (s - t)) - (r - t) = 2t$.
    This proves the existence.
\end{proof}

\begin{remark}
We make three remarks on this step.

First, on algorithmic implementation.
Once the existence of $(\xi_2,\xi_1,\xi_0)$ is established,
they can be found by standard linear algebraic techniques,
with time complexity polynomial in $q$.

Second, on necessity of the annihilator condition \eqref{eq: ann coker}.
We cannot lift $\mathbb{F}_2$-linear transformations in $\coker H_X$ 
when the annihilator condition is not satisfied.
In this case, the ring $R$ then acts on the anyon module $\coker H_X$ in a non-trivial way.
Therefore, an $\mathbb{F}_2$ basis for $\coker H_X$
cannot be extended to a minimal generating set for free module $R^s$.
For example, for 6.6.6 color code, an $\mathbb{F}_2$ basis for $\coker H_X$ can be $\{1,x\}$, but $s = 1$ so only one element (e.g., $\{1\}$) is enough for a minimal generating set for $R^s$.

Third, on the lift of $\mathbb{F}_2$-linear maps.
In the proof, we have used the fact that an $\mathbb{F}_2$-isomorphism can be lifted to an $R$-isomorphism.
This does not hold for a general residue field $R/I$ with maximal ideal $I$.
Consider an invertible $m\times m$ matrix $A$ over the field $R/I$.
Write the elements of $A$ using their representatives in $R$.
The necessary and sufficient condition for such an $A$ to be invertible over $R$ is that $\det A$ is a unit in $R$.
In our case, $R/J$ is the coefficient field, so the condition always holds.
A non-example is when one takes $I = (x+y, 1+x+x^2+x^3+x^4)$ such that $R/I = \mathbb{F}_{16}$. 
In this case $A = \diag(1, 1+x^2)$ is not liftable, while $A = \diag(1, 1 + x + x^3 + x^4)$ is liftable to $\diag(1, x^2)$.
This also demonstrates the necessity of the annihilator condition
from another viewpoint: it can be tricky if we transform
the anyon basis based on more complex field structures rather than
treating the anyons one by one.
\end{remark}

By Lemma~\ref{lem: clear in F2}, we may assume without loss of generality that $H_Z^\dagger \equiv \tilde H_Z^\dagger \pmod{J}$ and $H_X \equiv \tilde H_X \pmod{J}$: any isomorphism found under this assumption can be composed with $(\xi_2, \xi_1, \xi_0)$ to yield the maps of Theorem~\ref{thm: main}.
Under this assumption, both sequences become short exact with the common right-hand term $R^{p_X}\oplus JR^t$ with aligned generating elements ($X$-checks),
and our problem reduces to finding $R$-isomorphisms $(\phi_2, \phi_1)$ such that the following diagram commutes:
\begin{equation}
\begin{tikzcd}
	0 & {R^{r}} & {R^{q}} & {R^{p_X}\oplus JR^t} & 0 \\
	0 & {R^{r}} & {R^{q}} & {R^{p_X}\oplus JR^t} & 0
	\arrow[from=1-1, to=1-2]
	\arrow["{H_Z^\dagger}", from=1-2, to=1-3]
	\arrow["{H_X}", from=1-3, to=1-4]
	\arrow[from=1-4, to=1-5]
	\arrow[from=2-1, to=2-2]
	\arrow["\phi_2", from=2-2, to=1-2]
	\arrow["{\tilde{H}_Z^\dagger}"', from=2-2, to=2-3]
	\arrow["{\phi_1}", from=2-3, to=1-3]
	\arrow["{\tilde{H}_X}"', from=2-3, to=2-4]
	\arrow["{\id}", from=2-4, to=1-4]
	\arrow[from=2-4, to=2-5]
\end{tikzcd}
\label{eq: ses map}
\end{equation}

\subsection{Step 2. Existence}\label{sec:existence}

In this subsection, we explore the algebraic structure of our problem
through the lens of short exact sequences and the classification of module extensions.
This establishes the existence of the decoupling map in Theorem~\ref{thm: main}
and serves as a first step towards the algorithm that computes it.

First, recall some standard results from homological algebra (see e.g. Chapter XIV in Ref.~\cite{cartan1999homological}).
An extension of an $R$-module $A$ by an $R$-module $C$ is a short exact sequence $0\rightarrow C \xrightarrow{\psi} X \xrightarrow{\phi} A \rightarrow 0$.
Two extensions $X, X'$ are equivalent if there exists an $R$-isomorphism $k\colon X \rightarrow X'$ making the following diagram commute:
\begin{equation}
    \begin{tikzcd}
        & X \\
        C && A \\
        & {X'}
        \arrow["\phi", from=1-2, to=2-3]
        \arrow["k", from=1-2, to=3-2]
        \arrow["\psi", from=2-1, to=1-2]
        \arrow["{\psi'}"', from=2-1, to=3-2]
        \arrow["{\phi'}"', from=3-2, to=2-3]
    \end{tikzcd}
\end{equation}
Furthermore, classification of extensions under this equivalence is captured by the Ext functor.
Let $E(A,C)$ denote the set of equivalence classes of extensions of $A$ by $C$.
There is a bijection $\Theta\colon E(A,C) \to \Ext^1(A,C)$ defined by $\Theta(E) = \delta_E(\id)$,
where $\delta_E\colon \Hom_R(C,C) \rightarrow \Ext^1(A,C)$ is the connecting homomorphism obtained by applying $\Hom_R(-,C)$ to the extension $E$,
and $\id \in \Hom_R(C,C)$ is the identity map.

The existence of isomorphisms $(\phi_2,\phi_1)$ can be reduced to a problem of equivalence of extensions by absorbing (the unfixed) $\phi_2$ into the input chain.
Consider the chain complex
\begin{equation}
    0 \rightarrow R^r \xrightarrow{H_Z^\dagger \circ \phi_2} R^q \xrightarrow{H_X} R^{p_X}\oplus J R^t \rightarrow 0.
    \label{eq: pushout chain}
\end{equation}
If there exists some $\phi_2$ such that 
this chain complex is equivalent to the standard chain complex (second row in Eq.~\eqref{eq: ses map}),
i.e., if it has the same representative in $\Ext^1(R^{p_X}\oplus JR^t, R^r)\cong \Hom_{\mathbb{F}_2}(\mathbb{F}_2^t, \mathbb{F}_2^r)$
as the standard one,
then the required $(\phi_2,\phi_1)$ exists.
Therefore, we can establish the following existence result
by proving that the orbit of the representatives of
\eqref{eq: pushout chain} under all possible invertible $\phi_2$ (called ``pushout'') 
contains the representative of the standard chain.

\begin{lemma}\label{lemma: existence}
    There exist $R$-isomorphisms $(\phi_2,\phi_1)$ such that the diagram~(\ref{eq: ses map}) commutes.
\end{lemma}

\begin{proof}
    From standard homological algebra we have 
    $\Ext^1(R^{p_X}\oplus JR^t, R^r)\cong \Hom_{\mathbb{F}_2}(\mathbb{F}_2^t, \mathbb{F}_2^r)$,
    so that we can write the representatives as $r \times t$ $\mathbb{F}_2$-matrices.
    Furthermore, let the representative of the first row in \eqref{eq: ses map} be $\theta$, 
    then the representative of the pushed out chain \eqref{eq: pushout chain} is $(\phi_2^{-1}\bmod J) \theta$.
    Also denote the representative of the standard chain as $\tilde\theta$.
    We will prove that both $\theta$ and $\tilde\theta$ have rank (as $\mathbb{F}_2$-matrices) $t$.
    Therefore, there exists an invertible $r\times r$ $\mathbb{F}_2$-matrix $\alpha$ such that $\alpha \theta = \tilde\theta$ (note that $r \geq t$).
    Then taking $\phi_2 = \alpha^{-1}$ finishes the proof.

    That both $\theta$ and $\tilde\theta$ have full rank $t$ is a consequence of the shape of the middle module $R^q$.
    Consider the following building blocks:
    \begin{itemize}
        \item The trivial extensions $E(0,R)$ and $E(R,0)$, each with middle module $R$.
        \item The extensions in $E(J, R)$, classified by $\Ext^1(J, R)\cong \Hom_{\mathbb{F}_2}(\mathbb{F}_2, \mathbb{F}_2)\cong \mathbb{F}_2$.
        There are exactly two classes:
        the split extension with middle module $R\oplus J$, and the non-split extension with middle module $R^2$;
        the latter is the toric code sequence (see Eq.~\eqref{eq: toric code check matrix}).
    \end{itemize}
    Taking the direct sum of $p_X$ copies of $E(0,R)$, $p_Z$ copies of $E(R,0)$, $\Delta$ split copies of $E(J,R)$, and $t-\Delta$ non-split (toric code) copies yields the reference extension
    \begin{equation}
        E_\Delta\colon 0 \rightarrow R^r \rightarrow R^{q-\Delta}\oplus JR^\Delta \rightarrow R^{p_X}\oplus JR^t \rightarrow 0,
    \end{equation}
    where $q = p_X + p_Z + 2t$.
    By additivity of $\Ext^1$, the representative $\Theta(E_\Delta) \in \Hom_{\mathbb{F}_2}(\mathbb{F}_2^t, \mathbb{F}_2^r)$ has rank $t - \Delta$, contributed entirely by the non-split (toric code) summands.
    Based on these reference extensions, 
    pushing out in $R^r$ and pulling back in the direct summand $J R^t$ by invertible $\mathbb{F}_2$-matrices (intuitively, changes of ``basis'' in $R^r$ and $J R^t$) act on the representatives by
    left and right multiplication by some $\mathbb{F}_2$-invertible matrices
    (intuitively, change of basis in $\mathbb{F}_2^r$ and $\mathbb{F}_2^t$),
    which traverses all $r\times t$ matrices of the same rank.
    Therefore, every extension is equivalent to some reference extension up to pushouts and pullbacks,
    and that the middle module is $R^{q-\Delta}\oplus JR^\Delta$
    if and only if the representative has rank $t-\Delta$.
    This proves that both $\theta$ and $\tilde\theta$ have rank $t$.
\end{proof}

\begin{remark}
    For full rank $t$, pushouts alone (without pullbacks) suffice to traverse the representatives.
    This is why we can fix the zero-level map to $\id$.
\end{remark}

\subsection{Step 3. Construction and algorithm}\label{sec:c&a}

In this subsection, we proceed from existence to construction.
Our approach follows the ``directly solve the matrix'' route.
There we noted that this strategy does not work naively: the maps have nontrivial kernels, and these kernels are not direct summands.
However, we can take a good ansatz of $\phi_1$ and then correct it to an isomorphism by exploiting more algebraic properties of our problem.

The first step is to apply the comparison theorem for projective resolutions (see Ref.~\cite{weibelIntroductionHomologicalAlgebra1994}, Comparison Theorem~2.2.6) to our setting
to reveal more constraints on $\phi_2$.

\begin{lemma}\label{lemma: comparison}
Let
\begin{equation}
    0 \rightarrow R^r \xrightarrow{H_Z^\dagger} R^q \xrightarrow{H_X} R^{p_X}\oplus JR^t \rightarrow 0
\end{equation}
be a short exact sequence with $H_X \equiv \tilde{H}_X \pmod{J}$
and $H_Z^\dagger \equiv \tilde{H}_Z^\dagger \pmod{J}$,
and let $(\phi_2, \phi_1)$ be an isomorphism between this sequence
and the standard sequence,
as guaranteed by Lemma~\ref{lemma: existence}.
Then the reduction mod $J$ of the last $t$ columns of $\phi_2$
is independent of the choice of isomorphism
and has the form
\begin{equation}
    (\phi_2)_{[:, p_Z+1:r]} \equiv
    \begin{pmatrix}
        0_{p_Z \times t} \\
        B_t
    \end{pmatrix} \pmod{J},
\end{equation}
where $B$ is a fixed $t\times t$ invertible matrix over $\mathbb{F}_2$.
\end{lemma}

\begin{proof}
    Since $\phi_2$ acts on the representative in $E(R^{p_X}\oplus JR^t, R^r)$
    by left multiplication by $\phi_2^{-1} \bmod J$,
    there exists an $\mathbb{F}_2$-matrix $(\phi_2)_0$
    that is an $R$-isomorphism and makes the diagram \eqref{eq: ses map} commute with some isomorphism $\phi_1$,
    according to Lemma~\ref{lemma: existence}.
    We first prove that such a $(\phi_2)_0$ can have
    the following block diagonal form:
    \begin{equation}
        (\phi_2)_0 = \begin{pmatrix}
            A_{p_Z} & 0_{p_Z\times t} \\ C_{t\times p_Z} & B_t
        \end{pmatrix},
        \label{eq: F2 diag rho}
    \end{equation}
    with $A_{p_Z}, B_t$ being invertible $\mathbb{F}_2$-matrices.
    In fact, for such $(\phi_2)_0$, reducing diagram~(\ref{eq: ses map}) modulo $J$ gives
    $(H_Z^\dagger)_\ast (\phi_2)_0 = (\phi_1)_\ast (\tilde H_Z^\dagger)_\ast$.
    By assumption $(H_Z^\dagger)_\ast = (\tilde H_Z^\dagger)_\ast$, and the standard form~\eqref{eq: goal matrices} gives
    $\ker (\tilde H_Z^\dagger)_\ast = \pi_{[p_Z+1:r]} \mathbb{F}_2^r$.
    For any $v \in \pi_{[p_Z+1:r]} \mathbb{F}_2^r$, we have $(\tilde H_Z^\dagger)_\ast v = 0$,
    so $(H_Z^\dagger)_\ast (\phi_2)_0 v = (\phi_1)_\ast (\tilde H_Z^\dagger)_\ast v = 0$,
    i.e., $(\phi_2)_0 v \in \ker (H_Z^\dagger)_\ast = \pi_{[p_Z+1:r]} \mathbb{F}_2^r$.
    Hence $\pi_{[p_Z+1:r]} \mathbb{F}_2^r$ is invariant under $(\phi_2)_0$.
    This proves the diagonal form in Eq.~\eqref{eq: F2 diag rho}.
    Invertibility of $(\phi_2)_0$ thus implies invertibility of $A$ and $B$.
    
    We then prove that the last $t$ columns for any $\phi_2$
    are the same as $(\phi_2)_0$ modulo $J$.
    Since $R^r$ and $R^q$ are free (hence projective) $R$-modules,
    each row in the diagram below is a projective resolution
    of $R^{p_X}\oplus JR^t$,
    and $(\phi_2, \phi_1)$ is a chain map lifting $\id$ on the cokernel:
    \begin{equation}
        \begin{tikzcd}
        	0 & {R^{r}} & {R^{q}} & {R^{p_X}\oplus JR^t} & 0 \\
        	0 & {R^{r}} & {R^{q}} & {R^{p_X}\oplus JR^t} & 0
        	\arrow[from=1-1, to=1-2]
        	\arrow["{H_Z^\dagger}", from=1-2, to=1-3]
        	\arrow["{H_X}", from=1-3, to=1-4]
        	\arrow[from=1-4, to=1-5]
        	\arrow[from=2-1, to=2-2]
        	\arrow["\phi_2", from=2-2, to=1-2]
        	\arrow["{\tilde{H}_Z^\dagger}"', from=2-2, to=2-3]
        	\arrow["{\phi_1}", from=2-3, to=1-3]
        	\arrow["{\eta}", from=2-3, to=1-2, dashed]
        	\arrow["{\tilde{H}_X}"', from=2-3, to=2-4]
		\arrow["{\id}", from=2-4, to=1-4]
        	\arrow[from=2-4, to=2-5]
        \end{tikzcd}
        \label{eq: comparison}
    \end{equation}
    By the comparison theorem, such lifts are unique up to chain homotopy:
    any choice of $\phi_2$ satisfies
    $\phi_2 = (\phi_2)_0 + \eta\, \tilde{H}_Z^\dagger$
    for some $R$-map $\eta\colon R^q \to R^r$.
    By the block form of $\tilde{H}_Z^\dagger$ \eqref{eq: goal matrices},
    its last $t$ columns lie in $JR^q$,
    so $\eta\, (\tilde{H}_Z^\dagger)_{[:,\,p_Z+1:r]} \subseteq JR^r$
    for any $\eta$.
    Hence the last $t$ columns of $\phi_2$ are uniquely determined modulo $J$
    independently of the choice of lift.
    This proves the lemma.
\end{proof}

The preceding lemma suggests a strategy: 
if we start with a sufficiently structured initial ansatz for $\phi_1$,
the corresponding $\phi_2$ will be close to the block-diagonal form of Eq.~\eqref{eq: F2 diag rho}
and can then be corrected to an isomorphism via the homotopy map $\eta$. 

\begin{lemma}\label{lemma: key for alg}
    Let
    \begin{equation}
        0 \rightarrow R^r \xrightarrow{H_Z^\dagger} R^q \xrightarrow{H_X} R^{p_X}\oplus JR^t \rightarrow 0
    \end{equation}
    be a short exact sequence with $H_X \equiv \tilde{H}_X \pmod{J}$ and $H_Z^\dagger \equiv \tilde{H}_Z^\dagger \pmod{J}$.
    Then there exists a pair of homomorphisms $(\phi_2, \phi_1)$ such that the diagram~\eqref{eq: comparison} commutes (ignoring the dashed arrow), with the following properties:
    \begin{itemize}
        \item The $p_X+1$ to $p_X+p_Z$ columns of $\phi_1$ coincide with the first $p_Z$ columns of $H_Z^\dagger$.
        \item The matrix $\phi_2$ has the block form
        \begin{equation}
            \phi_2 = \begin{pmatrix}
                I_{p_Z} & C_{p_Z\times t} \\ 0_{t\times p_Z} & B_t + D_t
            \end{pmatrix},
            \label{eq: initial form rho}
        \end{equation}
        where $B_t$ is an invertible $\mathbb{F}_2$-matrix and all entries of $C$ and $D$ lie in $J$.
    \end{itemize}
    Furthermore, there exists $\eta \in \Hom(R^q, R^r)$ such that $\phi_2 + \eta \tilde{H}_Z^\dagger$ is an $R$-isomorphism.
\end{lemma}

\begin{proof}
    For the form of $\phi_1$, the right square of diagram~\eqref{eq: comparison} requires $H_X \phi_1 = \tilde{H}_X$.
    Columns $p_X+1$ through $p_X+p_Z$ of $\tilde{H}_X$ vanish; setting the corresponding columns of $\phi_1$ equal to the first $p_Z$ columns of $H_Z^\dagger$ satisfies this, since $H_X H_Z^\dagger = 0$.
    Each remaining column $q_i$ must satisfy $H_X q_i = (\tilde{H}_X)_{[:,i]}$, which is solvable because $\im H_X = \im \tilde{H}_X$, and can be computed efficiently via Gr\"obner bases for submodules.

    For the form of $\phi_2$, the left square requires $H_Z^\dagger \phi_2 = \phi_1 \tilde{H}_Z^\dagger$.
    Our choice of the $p_Z$ columns $[p_X+1:p_X+p_Z]$ of $\phi_1$, together with the injectivity of $H_Z^\dagger$, forces the first $p_Z$ columns of $\phi_2$ to be as in Eq.~\eqref{eq: initial form rho}.
    The form of the last $t$ columns of $\phi_2$ is ensured by
    Lemma~\ref{lemma: comparison}.

    For the existence of $\eta$, since every entry of $D_t$ lies in $J = (x-1, y-1)$, the equation
    \begin{equation}
        \begin{pmatrix}
            0_{p_Z} & \\ & -D_t
        \end{pmatrix} =
        \begin{pmatrix}
            0_{p_Z\times p_Z} & 0_{p_Z\times p_X} & 0_{p_Z\times 2t} \\
            0_{t\times p_Z} & 0_{t\times p_X} & E_{t\times 2t}
        \end{pmatrix}
        \begin{pmatrix}
            0_{p_X\times p_Z} & 0_{p_X\times t} \\ 
            I_{p_Z} & 0_{p_Z\times t} \\ 
            0_{t\times p_Z} & (1-y)I_{t} \\ 
            0_{t\times p_Z} & (x-1)I_{t} \\
        \end{pmatrix}
    \end{equation}
    has a solution $E$: it reduces to expressing each entry of $D_t$ as an $R$-linear combination of $x-1$ and $y-1$.
    Taking $\eta$ to be the left factor, we obtain
    \begin{equation}
        \phi_2 + \eta \tilde{H}_Z^\dagger = \begin{pmatrix}
            I_{p_Z} & C \\ 0 & B_t
        \end{pmatrix},
    \end{equation}
    which is upper block-triangular with invertible diagonal blocks, hence an $R$-isomorphism.
\end{proof}

We are now fully prepared to prove the main theorem.

\begin{proof}[Proof of Theorem~\ref{thm: main}]
    The number of toric-code sectors is $t = \dim_{\mathbb{F}_2} \coker H_X$, giving $p_Z = r - t$ and $p_X = s - t$.
    Lemma~\ref{lem: clear in F2} reduces the problem to the case $H_Z^\dagger \equiv \tilde{H}_Z^\dagger \pmod{J}$ and $H_X \equiv \tilde{H}_X \pmod{J}$, where both chain complexes become short exact sequences with the common right-hand term $R^{p_X} \oplus JR^t$.
    Lemma~\ref{lemma: existence} guarantees the existence of $R$-isomorphisms $(\phi_2, \phi_1)$ making diagram~\eqref{eq: ses map} commute, and Lemma~\ref{lemma: key for alg} shows how to construct them: one solves for an initial pair and corrects $\phi_2$ to an isomorphism via a chain homotopy, after which the five lemma forces $\phi_1$ to be an isomorphism as well.
    Algorithm~\ref{alg: constructing isomorphism} implements the procedure proposed in Lemma~\ref{lemma: key for alg}.
    It works in time polynomial in $q$ since every step is either Gaussian elimination over $\mathbb{F}_2$ or a module Gr\"obner basis computation in the four-variable polynomial lift of the Laurent ring.
    See Eq.~\eqref{eq:algo-locality-time} for the resulting time-complexity and locality bound.
\end{proof}

\section{Examples and numerical results}
\label{sec: examples}

In this section, we present examples and numerical results
of our decoupling algorithm.
We detail the computation of the supercell and the decoupling map
for the 6.6.6 and 4.8.8 color codes,
providing a pedagogical demonstration of the main results.
For convenience, we ignore the difference between $\pm 1$ in $\mathbb{F}_2$ in this section.
We also present numerical results
for a large number of bivariate bicycle codes,
demonstrating the validity of the algorithm.
The source code and data used to reproduce the numerical results are available
at \url{https://github.com/nzy1997/topological-css-code-decoupling}.

\subsection{The 6.6.6 color code}

\begin{figure}
    \centering
    \includegraphics[width=0.3\textwidth]{fig/666cc_basis_antipode.pdf}
    \hfill
    \includegraphics[width=0.3\textwidth]{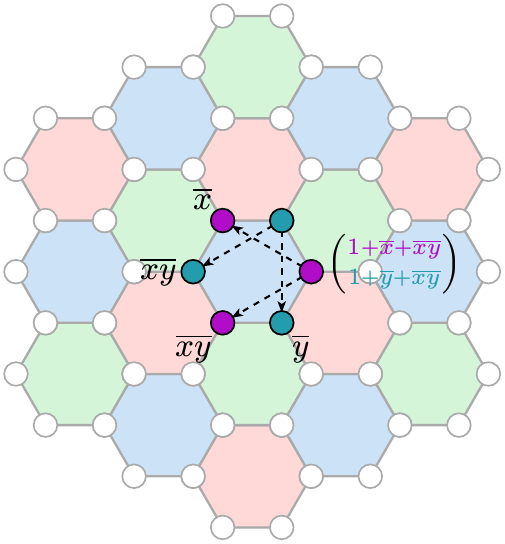}
    \hfill
    \includegraphics[width=0.3\textwidth]{fig/666cc_anyon.pdf}
    \caption{The 6.6.6 color code, with $q = 2$, $r = s = 1$.
    As shown in the first panel, there are two qubits per unit cell: the purple one is denoted by $(1,0)^\top$, while the green one is denoted by $(0,1)^\top$.
    The only $X$-type stabilizer generator is the face operator $(1+\bar{x}+\bar{xy},1+\bar{y}+\bar{xy})^\top$, and similarly for the $Z$-type stabilizer. For the 6.6.6 color code, the anyons respect the translation symmetry generated by $x^3$ and $x^2y$, as shown in the last panel.
    }
    \label{fig:666cc}
\end{figure}

The 6.6.6 color code is defined on a hexagonal lattice. In the basis shown in Fig.~\ref{fig:666cc}, the check matrix is given by:
\begin{equation}
    H_X = \begin{pmatrix}
        1+x+xy & 1+y+xy
    \end{pmatrix}.
    \label{eq:666cc_H_x}
\end{equation}
Following the procedure described in Section \ref{subsec: review wsb} and Algorithm~\ref{alg: constructing isomorphism}, we have:
\begin{enumerate}
    \item The Gr\"obner basis for ideal $(1+x+xy,1+y+xy)$ is $\{1+x+x^2,x+y,1+x+u,1+y+v\}$ (we lift the ideal $(1+x+xy,1+y+xy)$ in $R=\mathbb{F}_2[x^{\pm1},y^{\pm1}]$ to the ideal $(1+x+xy,1+y+xy,1+xu,1+yv)$ in $P=\mathbb{F}_2[x,u,y,v]$).
    \item The monomial basis for the quotient $P/(1+x+xy,1+y+xy,1+xu,1+yv)$ is $\{1,x\}$.
    \item To compute the representation matrix, we have:
    \begin{equation*}
        \begin{aligned}
            \hat{T}_x(1) &= x,\\
            \hat{T}_x(x) &= x^2=1+x+(1+x+x^2)\equiv 1+x,\\
            \hat{T}_y(1) &= y=x+(x+y)\equiv x,\\
            \hat{T}_y(x) &= xy=x^2+x(x+y)=1+x+(1+x+x^2)+x(x+y)\equiv 1+x.
        \end{aligned}
    \end{equation*}
    Therefore,
    \begin{equation}
        T_x = \begin{pmatrix}
            0 & 1\\
            1 & 1
        \end{pmatrix},
        \qquad
        T_y = \begin{pmatrix}
            0 & 1\\
            1 & 1
        \end{pmatrix}.
    \end{equation}
    \item It is easy to check that $T_x=T_y$ and $T_x^3=1$. Thus the superlattice is generated by $x^3, x^2y$.
    \item The coarse-grained check matrices are:
    \begingroup
    \setlength{\arraycolsep}{4pt}
    \begin{equation}
        H_X=
        \begin{pmatrix}
            1 & 1 & \bar{x} & 1 & 1 & \bar{y}\\
            y & 1 & 1 & x & 1 & 1\\
            xy & y & 1 & xy & x & 1
        \end{pmatrix},
        \qquad
        H_Z=
        \begin{pmatrix}
            1 & \bar{x} & \bar{x}\bar{y} & 1 & \bar{y} & \bar{x}\bar{y}\\
            1 & 1 & \bar{x} & 1 & 1 & \bar{y}\\
            y & 1 & 1 & x & 1 & 1
        \end{pmatrix}.
    \end{equation}
    \endgroup
    
    The corresponding standard forms are:
    \begingroup
    \setlength{\arraycolsep}{4pt}
    \begin{equation}
        \tilde H_X=
        \begin{pmatrix}
            1 & 0 & 0 & 0 & 0 & 0\\
            0 & 0 & x+1 & y+1 & 0 & 0\\
            0 & 0 & 0 & 0 & x+1 & y+1
        \end{pmatrix},
        \qquad
        \tilde H_Z=
        \begin{pmatrix}
            0 & 1 & 0 & 0 & 0 & 0\\
            0 & 0 & 1+\bar{y} & 1+\bar{x} & 0 & 0\\
            0 & 0 & 0 & 0 & 1+\bar{y} & 1+\bar{x}
        \end{pmatrix}.
    \end{equation}
    \endgroup
    
     Running Algorithm~\ref{alg: constructing isomorphism} gives the transformation matrices $(\psi^{-1}_2, \psi^{-1}_1, \psi^{-1}_0)$ in Theorem~\ref{thm: main}:
    
    \begin{equation}
        \psi^{-1}_0=
        \begin{pmatrix}
            1 & 0 & 0\\
            1 & 1 & 0\\
            1 & 0 & 1
        \end{pmatrix},
        \psi^{-1}_1=
        \begin{pmatrix}
            1+\bar{y} & 1 & 1 & 0 & 1+\bar{y} & 1+\bar{y}\\
            x\bar{y}+1+\bar{y} & x & 1 & 1 & x\bar{y}+1+\bar{y} & x\bar{y}+\bar{y}\\
            x\bar{y} & xy & 0 & 0 & x\bar{y} & x\bar{y}\\
            1 & 1 & 1 & 0 & 1 & 1\\
            1+\bar{y} & y & 0 & 0 & \bar{y} & 1+\bar{y}\\
            x+y & xy & y & y & x+y & x+y
        \end{pmatrix},
        \psi^{-1}_2=
        \begin{pmatrix}
            1 & 1 & 1+x\bar{y}\\
            0 & y & x+x\bar{y}+y\\
            0 & 0 & x
        \end{pmatrix}.
    \end{equation}
\end{enumerate}
\subsection{The 4.8.8 color code}

\begin{figure}
    \centering
    \includegraphics[width=0.3\textwidth]{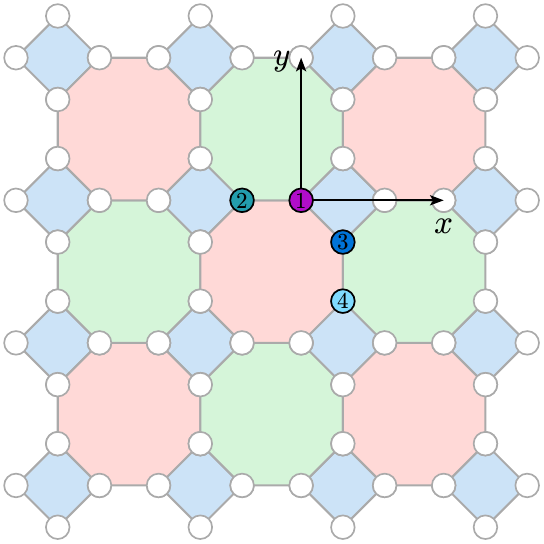}
    \hfill
    \includegraphics[width=0.3\textwidth]{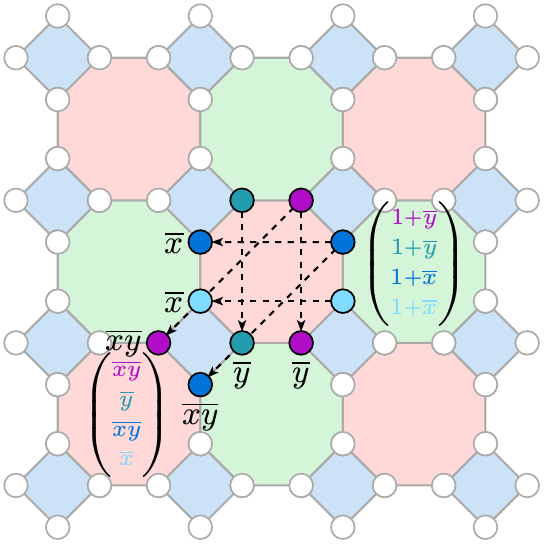}
    \hfill
    \includegraphics[width=0.3\textwidth]{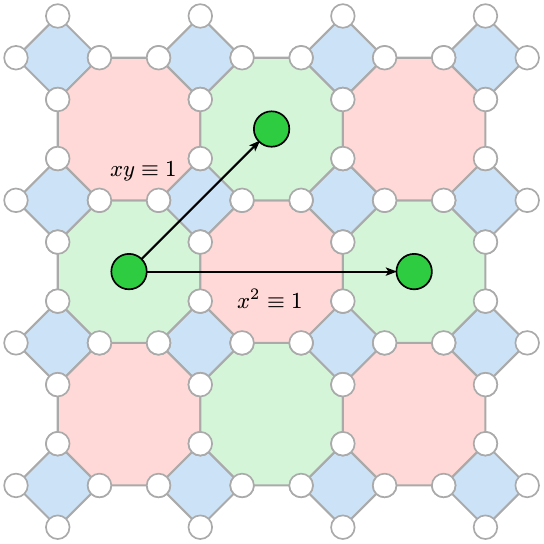}
    \caption{The 4.8.8 color code, with $q = 4$, $r = s = 2$.
    As shown in the first panel, there are four qubits per unit cell, denoted by $e_i$, the vector with a $1$ in the $i$-th position and $0$s elsewhere.
    There are two different $X$-type stabilizer generators, denoted by $(1+\bar{y},1+\bar{y},1+\bar{x},1+\bar{x})^\top$ and $(\bar{xy},\bar{y},\bar{xy},\bar{x})^\top$, respectively (see the second panel), and similarly for the $Z$-type stabilizer. For the 4.8.8 color code, the anyons respect the translation symmetry generated by $x^2$ and $xy$, as shown in the last panel.
    }
    \label{fig:488cc}
\end{figure}

The 4.8.8 color code is defined on a square-octagon lattice. In the basis shown in Fig.~\ref{fig:488cc}, the check matrix is given by:
\begin{equation}
    H_X = \begin{pmatrix}
        1+y & 1+y & 1+x & 1+x\\
        xy & y & xy & x
    \end{pmatrix}
\end{equation}
Following the same procedure, we have:
\begin{enumerate}
    \item The Gr\"obner basis for submodule $\im{H_X}$ is given by:
    \begin{equation*}
        \left\{
        \begin{pmatrix}
            1+x\\1
        \end{pmatrix},    
        \begin{pmatrix}
            0\\1+x
        \end{pmatrix},    
        \begin{pmatrix}
            1+y\\1
        \end{pmatrix},    
        \begin{pmatrix}
            0\\1+y
        \end{pmatrix},    
        \begin{pmatrix}
            1+u\\1
        \end{pmatrix},    
        \begin{pmatrix}
            0\\1+u
        \end{pmatrix},    
        \begin{pmatrix}
            1+v\\1
        \end{pmatrix},    
        \begin{pmatrix}
            0\\1+v
        \end{pmatrix}
        \right\}
    \end{equation*}
    \item The monomial basis for the quotient $(P/I)^2/\im{H_X}$ is:
    \begin{equation*}
        \left\{
        \begin{pmatrix}
            1\\0 
        \end{pmatrix},
        \begin{pmatrix}
            0\\1 
        \end{pmatrix}
        \right\}
    \end{equation*}
    \item To compute the representation matrix, we have:
    \begin{equation*}
        \begin{aligned}
            \hat{T}_x\begin{pmatrix}
                1\\0
            \end{pmatrix} &= \begin{pmatrix}
                x\\0
            \end{pmatrix} = \begin{pmatrix}
                1\\0 
            \end{pmatrix} + \begin{pmatrix}
                0\\1 
            \end{pmatrix} + \begin{pmatrix}
                1+x\\1 
            \end{pmatrix} \equiv \begin{pmatrix}
                1\\0
            \end{pmatrix} + \begin{pmatrix}
                0\\1 
            \end{pmatrix}, \\\hat{T}_x\begin{pmatrix}
                0\\1
            \end{pmatrix} &= \begin{pmatrix}
                0\\x
            \end{pmatrix} = \begin{pmatrix}
                0\\1 
            \end{pmatrix} + \begin{pmatrix}
                0\\1+x 
            \end{pmatrix} \equiv \begin{pmatrix}
                0\\1
            \end{pmatrix},
            \\
            \hat{T}_y\begin{pmatrix}
                1\\0
            \end{pmatrix} &= \begin{pmatrix}
                y\\0
            \end{pmatrix} = \begin{pmatrix}
                1\\0 
            \end{pmatrix} + \begin{pmatrix}
                0\\1 
            \end{pmatrix} + \begin{pmatrix}
                1+y\\1 
            \end{pmatrix} \equiv \begin{pmatrix}
                1\\0
            \end{pmatrix} + \begin{pmatrix}
                0\\1 
            \end{pmatrix},
            \\
            \hat{T}_y\begin{pmatrix}
                0\\1
            \end{pmatrix} &= \begin{pmatrix}
                0\\y
            \end{pmatrix} = \begin{pmatrix}
                0\\1 
            \end{pmatrix} + \begin{pmatrix}
                0\\1+y 
            \end{pmatrix} \equiv \begin{pmatrix}
                0\\1
            \end{pmatrix}.
        \end{aligned}
    \end{equation*}
    \item Therefore,
    \begin{equation}
        T_x = \begin{pmatrix}
            1 & 0\\
            1 & 1
        \end{pmatrix},
        \qquad
        T_y = \begin{pmatrix}
            1 & 0\\
            1 & 1
        \end{pmatrix}.
    \end{equation}
    It is easy to check that $T_x=T_y$ and $T_x^2=1$. Therefore the superlattice is generated by $x^2, xy$.
    \item The coarse-grained check matrices are:
    \begingroup
    \setlength{\arraycolsep}{4pt}
    \begin{equation}
        H_X=
        \begin{pmatrix}
            1 & 1 & 1 & 1 & 1 & \bar{x}y & 1 & \bar{x}y\\
            x & 1 & x & 1 & y & 1 & y & 1\\
            y & 0 & 0 & 1 & y & 0 & 0 & \bar{x}y\\
            0 & y & x & 0 & 0 & y & y & 0
        \end{pmatrix},
        \qquad
        H_Z=
        \begin{pmatrix}
            1 & 1 & 1 & 1 & 1 & \bar{x}y & 1 & \bar{x}y\\
            x & 1 & x & 1 & y & 1 & y & 1\\
            y & 0 & 0 & 1 & y & 0 & 0 & \bar{x}y\\
            0 & y & x & 0 & 0 & y & y & 0
        \end{pmatrix}.
    \end{equation}
    \endgroup
    
    The corresponding standard forms are given by:
    \begingroup
    \setlength{\arraycolsep}{4pt}
    \begin{equation}
        \tilde H_X=
        \begin{pmatrix}
            1 & 0 & 0 & 0 & 0 & 0 & 0 & 0\\
            0 & 1 & 0 & 0 & 0 & 0 & 0 & 0\\
            0 & 0 & 0 & 0 & x+1 & y+1 & 0 & 0\\
            0 & 0 & 0 & 0 & 0 & 0 & x+1 & y+1
        \end{pmatrix},
        \qquad
        \tilde H_Z=
        \begin{pmatrix}
            0 & 0 & 1 & 0 & 0 & 0 & 0 & 0\\
            0 & 0 & 0 & 1 & 0 & 0 & 0 & 0\\
            0 & 0 & 0 & 0 & 1+\bar{y} & 1+\bar{x} & 0 & 0\\
            0 & 0 & 0 & 0 & 0 & 0 & 1+\bar{y} & 1+\bar{x}
        \end{pmatrix}.
    \end{equation}
    \endgroup
    
    Running Algorithm~\ref{alg: constructing isomorphism} gives the transformation matrices $(\psi^{-1}_2, \psi^{-1}_1, \psi^{-1}_0)$ in Theorem~\ref{thm: main}:
    \begingroup
    \setlength{\arraycolsep}{4pt}
    \begin{equation}
    \begin{aligned}
        \psi^{-1}_0&=
        \begin{pmatrix}
            1 & 0 & 0 & 0\\
            1 & 0 & 1 & 0\\
            1 & 1 & 0 & 0\\
            0 & 1 & 0 & 1
        \end{pmatrix},
        \qquad
        \psi^{-1}_2=
        \begin{pmatrix}
            1 & 0 & y & y\\
            0 & 0 & 0 & x\\
            0 & 1 & y & 0\\
            0 & 0 & xy & 0
        \end{pmatrix},
        \\
        \psi^{-1}_1&=
        \begin{pmatrix}
            0 & 0 & 1 & \bar{y} & 1 & 0 & 1 & 0\\
            0 & 0 & 1 & 0 & 1 & 1 & 1 & 1\\
            0 & 0 & 1 & 0 & 0 & 0 & 1 & 0\\
            1 & 0 & 1 & 1 & 0 & 0 & 1 & 1\\
            0 & \bar{y} & 1 & \bar{y} & 1 & 0 & 1+\bar{y} & \bar{y}\\
            0 & 0 & x\bar{y} & 0 & 0 & 0 & 0 & 0\\
            0 & \bar{y} & 1 & 0 & 1 & 1 & 1+\bar{y} & \bar{y}\\
            0 & 0 & x\bar{y} & x\bar{y} & 0 & 0 & 0 & 0
        \end{pmatrix}.
    \end{aligned}
    \end{equation}
    \endgroup
\end{enumerate}

\subsection{More results on Bivariate Bicycle codes}
\label{sec:bb-more-results}

We benchmark Algorithm~\ref{alg: constructing isomorphism} on two collections of bivariate bicycle (BB) codes. For each instance, we choose a coarse-graining superlattice and compute the decoupling map to the standard form. The tables report the resulting standard-form data, running time, and maximal Laurent degree $\deg(\psi^{-1}_1)$ of the qubit map. We use $\deg(\psi^{-1}_1)$ as a proxy for locality, since it bounds the range over which the decoupling map can spread a local $X$ operator.

\begin{table}[htbp]
    \centering
    \begingroup
    \scriptsize
    \setlength{\tabcolsep}{2.6pt}
    \renewcommand{\arraystretch}{1.08}
    \begin{tabular*}{\textwidth}{@{\extracolsep{\fill}}llllrrrrr}
        \hline
        $a(x,y)$ & $b(x,y)$ & $L$ & superlattice generator & $q$ & $p_X (= p_Z)$ & $t$ & Time(s) & deg($\psi^{-1}_1$) \\
        \hline
        $x \cdot y$ & $x \cdot y$ & $3$ & $(x^{2}y, xy^{2})$ & 6 & 1 & 2 & 0.014 & 6 \\
        $x^{-1} \cdot y$ & $x \cdot y$ & $7$ & $(x^{3}y, x^{2}y^{3})$ & 14 & 4 & 3 & 0.143 & 9 \\
        $x^{2}$ & $x^{2}$ & $3$ & $(x^{2}y, xy^{2})$ & 6 & 1 & 2 & 0.013 & 6 \\
        $x^{-1}$ & $y^{-1}$ & $3$ & $(x^{3}, y^{3})$ & 18 & 5 & 4 & 0.054 & 3 \\
        $x \cdot y$ & $x \cdot y^{-1}$ & $7$ & $(x^{3}y^{2}, xy^{3})$ & 14 & 4 & 3 & 0.03 & 9 \\
        $x^{-1}$ & $x^{3} \cdot y^{2}$ & $3$ & $(x^{3}, y^{3})$ & 18 & 5 & 4 & 0.027 & 12 \\
        $y^{-2}$ & $x^{-2}$ & $21$ & $(x^{5}y^{2}, x^{2}y^{5})$ & 42 & 16 & 5 & 0.184 & 14 \\
        $y^{-2}$ & $x^{2}$ & $15$ & $(x^{3}y^{3}, y^{5})$ & 30 & 11 & 4 & 0.109 & 13 \\
        $x^{-1} \cdot y$ & $x^{-1} \cdot y^{-1}$ & $31$ & $(x^{5}y^{3}, x^{3}y^{8})$ & 62 & 26 & 5 & 0.514 & 11 \\
        $x^{-2} \cdot y^{-1}$ & $x^{2} \cdot y$ & $3$ & $(xy, y^{3})$ & 6 & 1 & 2 & 0.014 & 6 \\
        $x^{-1} \cdot y^{3}$ & $x^{3} \cdot y^{-1}$ & $12$ & $(x^{12}, y^{12})$ & 288 & 136 & 8 & 9.29 & 36 \\
        $x^{-2}$ & $x^{-2} \cdot y^{2}$ & $7$ & $(x^{7}, y^{7})$ & 98 & 43 & 6 & 0.785 & 21 \\
        $x^{-2} \cdot y$ & $x \cdot y^{-2}$ & $63$ & $(x^{15}y^{6}, x^{6}y^{15})$ & 378 & 181 & 8 & 26.518 & 63 \\
        $x^{-1} \cdot y^{2}$ & $x^{-2} \cdot y^{-1}$ & $217$ & $(x^{19}y^{10}, x^{3}y^{13})$ & 434 & 209 & 8 & 40.519 & 90 \\
        $x^{-3} \cdot y$ & $x^{3} \cdot y^{2}$ & $63$ & $(x^{21}, y^{63})$ & 2646 & 1312 & 11 & 17056.737 & 210 \\
        $x^{-3} \cdot y$ & $x^{-5}$ & $21$ & $(xy, y^{21})$ & 42 & 16 & 5 & 0.307 & 37 \\
        $x^{-2} \cdot y$ & $x \cdot y^{2}$ & $105$ & $(x^{9}y, x^{3}y^{12})$ & 210 & 98 & 7 & 5.353 & 35 \\
        $x^{-1} \cdot y^{-2}$ & $x \cdot y^{-1}$ & $63$ & $(x^{7}y^{3}, y^{9})$ & 126 & 57 & 6 & 1.311 & 13 \\
        $y^{2}$ & $x^{-4} \cdot y$ & $73$ & $(x^{14}y^{5}, x^{5}y^{7})$ & 146 & 64 & 9 & 3.254 & 45 \\
        $x^{-1} \cdot y^{2}$ & $y^{-4}$ & $105$ & $(x^{35}y^{7}, y^{21})$ & 1470 & 725 & 10 & 979.291 & 147 \\
        $y^{-4}$ & $x^{4}$ & $255$ & $(x^{15}y^{13}, y^{17})$ & 510 & 239 & 16 & 79.217 & 95 \\
        $x^{-4}$ & $x^{-3} \cdot y^{2}$ & $21$ & $(x^{21}, y^{21})$ & 882 & 431 & 10 & 110.643 & 105 \\
        $x^{-2} \cdot y^{-5}$ & $x^{-1} \cdot y^{-3}$ & $42$ & $(x^{6}y^{6}, x^{2}y^{16})$ & 168 & 77 & 7 & 3.115 & 42 \\
        $x^{-8} \cdot y^{-1}$ & $x^{5} \cdot y$ & $21$ & $(x^{4}y, x^{3}y^{6})$ & 42 & 16 & 5 & 0.228 & 30 \\
        $x \cdot y^{-5}$ & $x \cdot y^{4}$ & $186$ & $(x^{18}y^{42}, x^{10}y^{44})$ & 744 & 363 & 9 & 119.246 & 174 \\
        $x^{-1} \cdot y^{-1}$ & $x^{5}$ & $105$ & $(x^{15}y^{6}, x^{10}y^{11})$ & 210 & 98 & 7 & 6.276 & 63 \\
        $x \cdot y^{3}$ & $x^{2} \cdot y^{-2}$ & $63$ & $(x^{35}y^{7}, x^{7}y^{14})$ & 882 & 432 & 9 & 167.668 & 126 \\
        $x^{-1} \cdot y^{-2}$ & $x^{2} \cdot y^{-1}$ & $217$ & $(x^{13}y^{3}, x^{10}y^{19})$ & 434 & 209 & 8 & 29.811 & 58 \\
        $x^{-1} \cdot y^{3}$ & $x \cdot y^{3}$ & $186$ & $(x^{16}y^{10}, x^{14}y^{32})$ & 744 & 363 & 9 & 147.209 & 138 \\
        $x^{2} \cdot y^{2}$ & $x^{-4} \cdot y$ & $889$ & $(x^{101}y^{5}, x^{4}y^{9})$ & 1778 & 879 & 10 & 1459.305 & 225 \\
        $x^{-1} \cdot y^{3}$ & $x^{3}$ & $217$ & $(x^{17}y^{9}, x^{8}y^{17})$ & 434 & 209 & 8 & 128.724 & 78 \\
        \hline
    \end{tabular*}
    \caption{Numerical results on the BB codes with the form $f=1+x+a(x,y)$ and $g=1+y+b(x,y)$ taken from Ref.~\cite{liangGeneralizedToricCodes2025a}. Here $L$ is the minimal translation period such that $(x^L-1,y^L-1)\subseteq I$, the superlattice generator column lists the two generators of the coarse-grained translation superlattice, $q$ is the number of qubits per supercell, $p_X(=p_Z)$ is the common number of product-state sectors of each type in the standard form, $t$ is the number of toric-code sectors, and $\deg(\psi^{-1}_1)$ is the maximal Laurent degree among the entries of the decoupling matrix $\psi^{-1}_1$. Only runs with a computed decoupling map $\psi^{-1}_1$ are listed.}
    \label{tab:bb-more-results}
    \endgroup
\end{table}

Table~\ref{tab:bb-more-results} gives results for the benchmark family of Ref.~\cite{liangGeneralizedToricCodes2025a}, defined by $f=1+x+a(x,y)$ and $g=1+y+b(x,y)$. Table~\ref{tab:additional-bb-code-results} gives results for additional BB codes taken from Refs.~\cite{zhouBunnyCodesBroadening2026,liangTopologicalSubsystemBivariate2026,symonsSequencesBivariateBicycle2025,bravyiHighthresholdLowoverheadFaulttolerant2024}.
In both tables, we list only instances corresponding to topological codes on the infinite plane whose supercell size $q$ is not too large so that the result can be obtained within a reasonable time.

\begin{table}[htbp]
    \centering
    \begingroup
    \fontsize{6}{6.8}\selectfont
    \setlength{\tabcolsep}{0.8pt}
    \renewcommand{\arraystretch}{1.03}
    \begin{tabular*}{\textwidth}{@{\extracolsep{\fill}}llllrrrrr}
        \hline
        $f$ & $g$ & $L$ & superlattice generator & $q$ & $p_X (= p_Z)$ & $t$ & Time(s) & deg($\psi^{-1}_1$) \\
        \hline
        $x + 1$ & $x + y$ & $1$ & $(x, y)$ & 2 & 0 & 1 & 0.012 & 0 \\
        $x + y^{2}$ & $y^{2} + y$ & $1$ & $(x, y)$ & 2 & 0 & 1 & 0.106 & 2 \\
        $x^{2} \cdot y + x \cdot y + x$ & $x \cdot y + y^{2}$ & $3$ & $(x^{2}y, xy^{2})$ & 6 & 1 & 2 & 0.047 & 9 \\
        $x^{2} + x + 1$ & $x + y$ & $3$ & $(x^{2}y, xy^{2})$ & 6 & 1 & 2 & 0.017 & 6 \\
        $x^{2} \cdot y + x \cdot y + 1$ & $x \cdot y + x$ & $3$ & $(x^{3}, y)$ & 6 & 1 & 2 & 0.015 & 4 \\
        $x^{2} \cdot y^{2} + x + 1$ & $x \cdot y^{2} + x \cdot y$ & $3$ & $(x^{3}, y)$ & 6 & 1 & 2 & 0.015 & 5 \\
        $x^{3} \cdot y + x^{3} + x^{2} \cdot y + x \cdot y$ & $x^{3} \cdot y + y^{2}$ & $4$ & $(xy, y^{4})$ & 8 & 1 & 3 & 0.055 & 6 \\
        $x^{2} \cdot y^{3} + x + 1$ & $x \cdot y^{3} + x \cdot y^{2} + x \cdot y$ & $3$ & $(x^{3}, y^{3})$ & 18 & 5 & 4 & 0.059 & 9 \\
        $x + y$ & $y^{4} + y^{3} + y^{2} + y$ & $4$ & $(x^{2}y^{2}, xy^{3})$ & 8 & 1 & 3 & 0.022 & 8 \\
        $y + x^{-1} \cdot y^{2} + x^{-2} \cdot y^{2} + x^{-2} \cdot y$ & $x \cdot y + y^{2} + 1 + x^{-2}$ & $3$ & $(xy, y^{3})$ & 6 & 0 & 3 & 0.032 & 6 \\
        $x \cdot y^{2} + x + x^{-1} \cdot y^{3} + x^{-2} \cdot y^{4}$ & $x \cdot y^{3} + x \cdot y + y^{2} + 1$ & $40$ & $(x^{2}y^{2}, y^{40})$ & 160 & 71 & 9 & 5.215 & 48 \\
        $x^{3} \cdot y^{-2} + x^{2} \cdot y^{-1} + y^{2} + 1$ & $x^{2} \cdot y + x^{2} \cdot y^{-1} + x + x \cdot y^{-2}$ & $40$ & $(x^{2}y^{2}, y^{40})$ & 160 & 71 & 9 & 7.337 & 120 \\
        $x^{3} \cdot y + 1 + x^{-1} \cdot y^{3} + x^{-2} \cdot y^{4}$ & $x^{3} \cdot y^{3} + x \cdot y + y^{2} + x^{-2}$ & $24$ & $(x^{4}y^{4}, y^{24})$ & 192 & 76 & 20 & 9.685 & 56 \\
        $x^{3} + x + x^{-1} \cdot y^{3} + x^{-2} \cdot y^{4}$ & $x^{3} \cdot y + x^{2} + x \cdot y + 1$ & $56$ & $(x^{2}y^{2}, y^{56})$ & 224 & 101 & 11 & 13.287 & 120 \\
        $x^{3} + y^{2} + y$ & $x^{2} + x + y^{3}$ & $12$ & $(x^{12}, y^{12})$ & 288 & 136 & 8 & 7.019 & 24 \\
        $x^{27} + x^{6} + y$ & $x^{24} + x^{15} + 1$ & $93$ & $(x^{93}, y)$ & 186 & 78 & 15 & 4.034 & 280 \\
        $x^{9} + y^{2} + y$ & $x^{8} + x + 1$ & $63$ & $(x^{36}y^{27}, x^{27}y^{36})$ & 1134 & 551 & 16 & 1085.181 & 252 \\
        $x^{3} + y^{2} + y$ & $x^{2} + x + 1$ & $3$ & $(x^{3}, y^{3})$ & 18 & 5 & 4 & 0.026 & 3 \\
        $x^{3} \cdot y + x^{2} \cdot y^{2} + 1$ & $x^{3} \cdot y^{2} + x^{2} + 1$ & $105$ & $(x^{21}y^{3}, y^{5})$ & 210 & 98 & 7 & 5.337 & 53 \\
        $x^{3} \cdot y^{2} + x^{2} \cdot y^{3} + 1$ & $x^{3} \cdot y + x^{2} + 1$ & $217$ & $(x^{14}y^{23}, x^{7}y^{27})$ & 434 & 209 & 8 & 27.003 & 74 \\
        $x^{3} \cdot y^{2} + x^{2} + 1$ & $x^{3} + x^{2} \cdot y + 1$ & $7$ & $(x^{7}, y)$ & 14 & 4 & 3 & 0.045 & 17 \\
        $x^{3} \cdot y + x^{2} + 1$ & $x^{3} \cdot y^{2} + x^{2} + 1$ & $7$ & $(x^{7}, y)$ & 14 & 4 & 3 & 0.028 & 16 \\
        $x^{3} + x^{2} + y$ & $x^{3} + x^{2} + 1$ & $7$ & $(x^{7}, y)$ & 14 & 4 & 3 & 0.022 & 7 \\
        $x^{6} \cdot y^{2} + x^{3} + x \cdot y^{3} + 1$ & $x^{6} \cdot y + x^{5} \cdot y + x^{4} \cdot y + x^{3} \cdot y^{2}$ & $120$ & $(x^{8}y^{6}, y^{30})$ & 480 & 230 & 10 & 94.622 & 104 \\
        $x^{6} + x^{3} + x \cdot y + 1$ & $x^{6} \cdot y + x^{5} \cdot y + x^{4} \cdot y + x^{3}$ & $8$ & $(x^{8}, y^{4})$ & 64 & 25 & 7 & 0.303 & 20 \\
        $x^{5} \cdot y + x^{5} + x^{3} + 1$ & $x^{5} \cdot y + x^{3} + x^{2} \cdot y + y$ & $186$ & $(x^{18}y^{2}, x^{12}y^{22})$ & 744 & 362 & 10 & 92.757 & 74 \\
        $x^{5} \cdot y^{3} + x^{4} \cdot y^{4} + x^{3} \cdot y^{5} + y^{5}$ & $x^{5} \cdot y + x^{4} \cdot y^{3} + x^{3} \cdot y^{3} + y^{2}$ & $56$ & $(x^{28}y^{2}, y^{4})$ & 224 & 103 & 9 & 11.048 & 120 \\
        $x^{5} \cdot y + x^{4} + x^{3} \cdot y + y$ & $x^{5} \cdot y + x^{4} \cdot y + x^{3} \cdot y + 1$ & $56$ & $(x^{28}y^{2}, y^{4})$ & 224 & 103 & 9 & 5.941 & 64 \\
        $x^{3} + y^{7} + y^{2}$ & $x^{2} + x + y^{3}$ & $126$ & $(x^{12}y^{6}, x^{6}y^{66})$ & 1512 & 744 & 12 & 865.028 & 234 \\
        \hline
    \end{tabular*}
    \caption{Results for more BB codes in Refs.~\cite{zhouBunnyCodesBroadening2026,liangTopologicalSubsystemBivariate2026,symonsSequencesBivariateBicycle2025,bravyiHighthresholdLowoverheadFaulttolerant2024}. The first two columns give the defining BB polynomial pair $(f,g)$. The remaining columns have the same meanings as in Table~\ref{tab:bb-more-results}; in particular, $q$ is the number of qubits per supercell.
    }
    \label{tab:additional-bb-code-results}
    \endgroup
\end{table}

\begin{figure}[htbp]
    \centering
    \includegraphics[width=0.8\columnwidth]{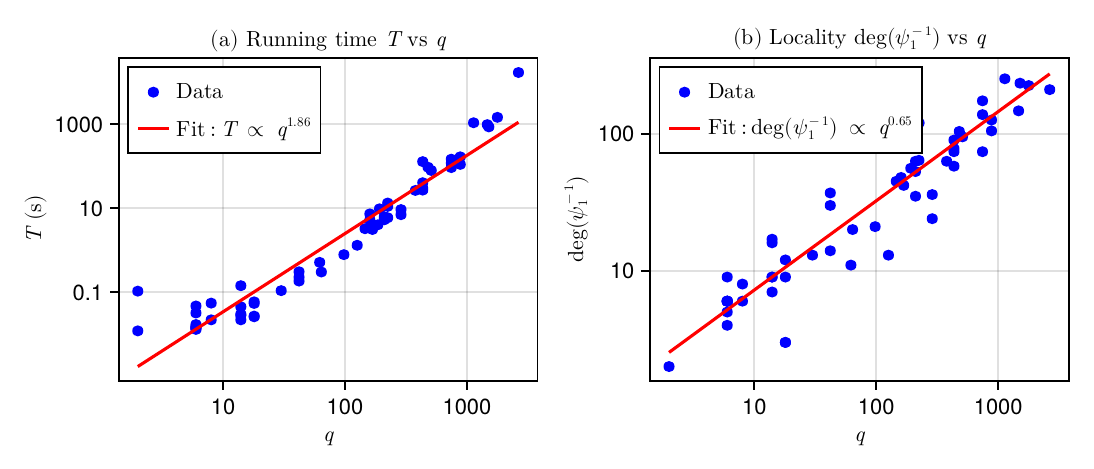}
    \caption{Empirical scaling of computation time and $\deg(\psi_1^{-1})$ versus qubit number per supercell $q$ for data from tables \ref{tab:bb-more-results} and \ref{tab:additional-bb-code-results}. The time fit uses all 60 entries, while the $\deg(\psi_1^{-1})$ fit uses the 59 non-zero entries. This scaling is much smaller than the pessimistic uniform module-theoretic worst-case bound in Eq.~\eqref{eq:algo-locality-time}.}
    \label{fig:area_time}
\end{figure}

Combining data from tables \ref{tab:bb-more-results} and \ref{tab:additional-bb-code-results} gives the empirical scaling shown in Fig.~\ref{fig:area_time}. A power-law fit over the combined data set gives
\begin{equation}
    T_{\mathrm{compute}} \propto q^{1.86},\qquad \deg(\psi_1^{-1})\propto q^{0.65}.
\end{equation}
This growth is far milder than the conservative worst-case estimate in Eq.~\eqref{eq:algo-locality-time}. Thus, while the uniform module-theoretic bound guarantees polynomial time in complete generality, the sparse structured BB code instances studied here appear much easier in practice. The locality estimate is also relevant for the unitary-decouple-based decoder in Sec.~\ref{sec:unitary-decouple-based-decoder}, where $\deg(\psi^{-1}_1)$ controls the possible spreading of a local physical fault across the virtual toric-code sectors.

The wall-clock timings reported in this section were measured on a Mac mini with an Apple M4 chip (10 CPU cores: 4 performance and 6 efficiency cores) and 32 GB of memory, running macOS 26.5.1 and Julia 1.12.5. 

\section{Corollaries and refinements}

In this section, we derive several consequences
of the decoupling algorithm developed above.
First, we show that the decoupling circuit
can always be made free of both ancillas and lattice shifts,
and comment on this result from the perspective of classification of quantum cellular automata.
Second, we show that the product states produced by the decoupling
can be reabsorbed into additional toric-code sectors,
resulting in a denser output.
In the literature, the product states in the output are sometimes also called ancillas.
Therefore, our discussion shows that we can achieve a clean decoupling that is ancilla-free on both the input and output sides.
Finally, we show that the maps constructed for the infinite plane
extend to codes with suitable periodic boundary conditions,
yielding a decoupling on the torus that preserves the logical algebra.

\subsection{Removing the shifts or ancillas}\label{sec: removing shifts}

In this subsection, we show that ancillas and lattice shifts are unnecessary for the decoupling.
A detailed discussion of the compilation problem is therefore necessary here.

By definition, the decomposition represented by the symplectic matrix $\diag(\psi_1,(\psi_1^\dagger)^{-1})$ is a 2D Clifford quantum cellular automaton (QCA).
Since this matrix is block diagonal, the QCA can be realized by lattice shifts (translations of superlattices) and a circuit containing only CNOT gates \cite{haahCliffordQuantumCellular2021}.
Such a lattice shift can be realized by local gates with the help of ancillas \cite{grossIndexTheoryOne2012},
which seems to indicate that ancillas are generally required to complete the decomposition without shifting the lattice.
However, we show that this is not the case via the following argument.
Given a circuit consisting of local Clifford gates and superlattice translations that decouples a 2D translation-invariant topological code,
we can always defer the translations to the last step.
Therefore, if we can construct a local unitary realization of translation on toric codes,
we succeed in showing that the circuit can be made free of both ancillas and shifts.

For a formal treatment, recall that for a pair of check matrices $(H_Z)_{r\times q}$ and $(H_X)_{s\times q}$,
the following matrix operations have direct physical interpretations.
\begin{itemize}
    \item Left-multiplying $H_X$ by an invertible $s\times s$ matrix $\psi_0$ to obtain $\psi_0 H_X$ amounts to a redefinition of stabilizer generators, as does left-multiplying $H_Z$ by an invertible $r\times r$ matrix $(\psi_2^\dagger)^{-1}$.
    \item Right-multiplying $H_X$ by a $q\times q$ 
    \textit{elementary matrix} $E_{i,j}(f(x,y))$ while simultaneously 
    right-multiplying $H_Z$ by 
    $(E^{\dagger}_{i,j}(f))^{-1} = E_{j,i}(-f(\overline{x},\overline{y}))$
    corresponds to a translation-invariant circuit of CNOT gates.
    Here $f$ is a polynomial,
    and $E_{i,j}(f) = I+e_{i,j}f(x,y)$ where $e_{i,j}$ is
    the $q\times q$ matrix whose $(i,j)$ component is 1 and 
    all other components are 0.
    If $f(x,y) = m(x,y)$ is a monomial, the corresponding circuit
    is a translation-invariant layer of CNOT gates,
    \begin{equation}
    \prod_{x^ay^b:(a,b)\in \mathbb{Z}^2} \text{CNOT}_{(x^a y^b,i),(m^{-1}x^a y^b,j)}.
    \end{equation}
    \item Right-multiplying both $H_X$ and $H_Z$ by the $q\times q$ matrix $E_{i}(m(x,y))$,
    where $m$ is a monomial and $E_i(m)$ is the diagonal matrix
    whose $i$'th diagonal component is $m$ and all others are 1.
    This corresponds to a lattice shift that shifts the $i$-th qubit in each unit cell by $m$.
\end{itemize}
Given an isomorphism $\psi_1$ from Algorithm~\ref{alg: constructing isomorphism},
we can decompose $\psi_1 =  E_1(\det \psi_1)\psi_1'$ such that $\det \psi_1' = 1$.
Note that $\det \psi_1$ is a monomial due to invertibility of $\psi_1$.
According to the Suslin stability theorem \cite{suslinStructureSpecialLinear1977, parkAlgorithmicProofSuslins1994},
the unimodular matrix $\psi_1'$ can be decomposed into a product of elementary matrices (matrices of the form $E_{i,j}(f(x,y))$ with $i\neq j$) whenever $q \geq 3$.
Therefore, if the basis qubit with label $1$ supports the product state
in the standard code,
the lattice shift by $\det \psi_1$ can be simply omitted.
This leads to the conclusion that if $q\geq 3$ and
there exist product states in the output standard code,
no lattice shift is needed for the decoupling.

The first requirement that $q\geq 3$ does not impair the conclusion
that decoupling can be made free of lattice shifts.
Recall that $q$ is the number of qubits per unit cell after coarse-graining.
For a coarse-grained code with $q < 3$, either $q=2$ so that it is already a toric code or a stack of product states,
or $q=1$ so that the code simply consists of product states.
In both cases the decoupling problem is trivial.

The second requirement that there exist product states
in the output standard code can be removed,
for even if the lattice shifts fall on
qubits that support the toric codes,
we can still find a constant-depth circuit to produce the same effect.
To see this, we only need to show that the effect of
any lattice shift on the toric-code stabilizer code can be reproduced by a constant-depth circuit.
In fact, recall that for toric codes,
\begin{equation}
\tilde H_X = (x-1, y-1),\quad \tilde H_Z^\dagger = \begin{pmatrix}
   1-y \\ x-1
\end{pmatrix}.
\end{equation}
By direct computation we have 
\begin{equation}
\begin{aligned}
    \tilde H_X E_1(x^ay^b)E_2(x^cy^d) &= x^a y^d \tilde H_X E_{1,2}\left((y-1)\frac{x^{c-a}-1}{x-1}\right) E_{2,1}\left(x^{a-c}(x-1)\frac{y^{b-d} - 1}{y-1}\right), \\ 
    \tilde H_Z E_1(x^ay^b)E_2(x^cy^d) &= x^c y^b \tilde H_Z E_{2,1}\left(-(\overline{y}-1)\frac{\overline{x}^{c-a}-1}{\overline{x}-1}\right) E_{1,2}\left(-\overline{x}^{a-c}(\overline{x}-1)\frac{\overline{y}^{b-d} - 1}{\overline{y}-1}\right).
\end{aligned}
\label{eq: shift unitary}
\end{equation}
Here $(y^k - 1)/(y-1)$ denotes the Laurent polynomial $1+y+\cdots + y^{k-1}$ for $k>0$, the polynomial $-y^{k}(1+{y} + \cdots + {y}^{-k-1})$ for $k<0$, and $0$ for $k = 0$; the expression $(x^k-1)/(x-1)$ is defined analogously.

\begin{figure}
    \centering
    \includegraphics[width=.25\textwidth]{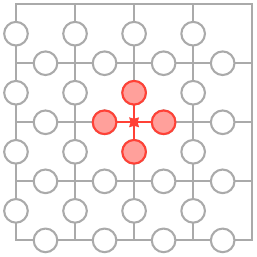}
    \hspace{2em}
    \includegraphics[width=.25\textwidth]{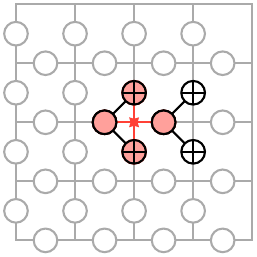}
    \hspace{2em}
    \includegraphics[width=.25\textwidth]{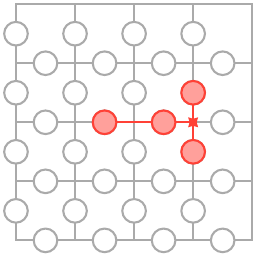}
    \caption{Illustration of simulating lattice shifts using a constant-depth circuit.
    The process shown here corresponds to Eq.~\eqref{eq: shift unitary} with $c = -1$, $a = b = d = 0$.}
    \label{fig:shift-demo}
\end{figure}

For example, consider the vertical-edge-shifted code
demonstrated in Fig.~\ref{fig:shift-demo} (c), with
\begin{equation}
    H_X = (x-1,\overline{x}(y-1)), H_Z = (1-\overline{y}, \overline{x}(\overline{x}-1)).
\end{equation}
We have $H_X = \tilde{H}_X E_2(\overline{x}) = \tilde{H}_X E_{1,2}(y-1)$ and $H_Z = \tilde{H}_ZE_{2,1}(1-\overline{y})$,
or equivalently 
\begin{equation}
    H_X  = \tilde{H}_XE_{1,2}(y-1),\quad H_Z  = \tilde{H}_ZE_{2,1}(-(\overline{y} - 1)).
\end{equation}
According to our discussion above, this means that
we can apply the translation-invariant CNOT circuit
demonstrated in Fig.~\ref{fig:shift-demo} (b)
to transform a standard toric code (Fig.~\ref{fig:shift-demo} (c))
to this vertical-edge-shifted one. 

It is known that a shift serves as an index for a 2D Clifford QCA, which cannot be replaced by any circuit without changing the QCA's behaviour.
Our construction does remove all shifts and ancillas from the decoupling procedure, seemingly a contradiction.
The resolution is that our map merely leaves the stabilizer code state invariant, rather than preserving the QCA itself.
Our result therefore does not violate the standard properties of QCA, but rather reinforces the intuition that the shift index measures the transfer of local information: 
since a topological code carries no local information, no shift is needed.

\subsection{Reabsorbing product states}

In many applications, it is preferable not to leave behind many product states.
For example, the celebrated unfolding of color codes produces two toric-code sectors on triangular lattices,
rather than two toric codes on square lattices with several product states~\cite{kubicaUnfoldingColorCode2015}.
A direct consequence of our algorithm is the ability to reabsorb these product states into toric codes.

Reabsorption is the reverse of decoupling product states
from a single toric-code sector.
Starting from a standard toric code on the square lattice,
we can choose a supercell containing any even number
$2\Lambda\geq 4$ of qubits.
Applying the decoupling algorithm to this coarse-grained code
yields one standard toric-code sector and $2\Lambda-2$ product-state sectors,
counted with respect to the coarse-grained lattice.
Alternatively, we can start from a toric code on the triangular lattice,
choose a supercell containing any multiple $3\Lambda\geq 3$ of qubits,
and apply the decoupling algorithm.
This yields one standard toric-code sector and $3\Lambda - 2$ product-state sectors.
Note that in both cases the number of qubits in the supercell
is at least $3$,
so the decoupling or reabsorption process is circuit-like.
We use the reverse of these two kinds of decoupling
to achieve reabsorption.

A meaningful reabsorption problem involves
at least $2$ toric-code sectors and $1$ product-state sector.
Let $p$ be the total number of product-state sectors.
For $p\neq 2$, there exist positive integers
$\Lambda_2$ and $\Lambda_3$ such that
$2\Lambda_2+3\Lambda_3=p+4$.
We can therefore reverse the corresponding decoupling processes
for square- and triangular-lattice toric codes with
$2\Lambda_2$ and $3\Lambda_3$ qubits per supercell, respectively, to reabsorb
all product states into two of the toric-code sectors while leaving
the others unchanged. For $p=2$, we instead reverse the square-lattice
decoupling with four qubits per supercell for one toric-code sector
and leave the other unchanged.

\begin{example}
Consider a single toric-code sector, coarse-grained with respect to $x^2$ and $y$. Applying the decoupling algorithm yields one toric-code sector together with two product-state sectors:
\begin{equation}
    \begin{aligned}
        H_X &= \begin{pmatrix}
            x + 1 & y + 1
        \end{pmatrix}, H_Z = \begin{pmatrix}
            \bar{y} + 1 & \bar{x} + 1
        \end{pmatrix}
        \xrightarrow{\text{coarse-graining}}
        \tilde H_X = \begin{pmatrix}
            1 & 1 & y + 1 & 0\\
            x & 1 & 0 & y + 1
        \end{pmatrix},
        \tilde H_Z = \begin{pmatrix}
            1+\bar{y} & 0 & 1 & \bar{x}\\
            0 & 1+\bar{y} & 1 & 1
        \end{pmatrix}
        \\
        &\xrightleftharpoons[
            \substack{\text{\quad Reabsorb product states\quad}\\\text{\quad into a toric code\quad}}
        ]{
            \text{\quad Decouple by Algorithm \ref{alg: constructing isomorphism}\quad}
        }
        \tilde H_X = \begin{pmatrix}
            1 & 0 & 0 & 0\\
            0 & 0 & x + 1 & y + 1
        \end{pmatrix},
        \tilde H_Z = \begin{pmatrix}
            0 & 1 & 0 & 0\\
            0 & 0 & \bar{y} + 1 & \bar{x} + 1
        \end{pmatrix}
    \end{aligned}
\end{equation}
\end{example}

\subsection{Periodic boundary conditions}\label{subsec: pbc}

The maps obtained for the infinite plane can be extended to codes with periodic boundary conditions (PBC).
Here we discuss the simplest case, in which the PBC is imposed after coarse-graining---that is, the translation vectors defining the torus correspond to translations that strictly preserve the anyons.

To impose periodic boundary conditions $x^a=1$, $y^b=1$, we tensor the exact sequence
\begin{equation}
    0 \rightarrow R^{r} \xrightarrow{H_Z^\dagger} R^{q} \xrightarrow{H_X}  R^{s} \rightarrow \mathbb{F}_2^t \rightarrow 0,
    \label{eq: free res chain}
\end{equation}
with $R/I$, where $I = (x^a-1,y^b-1)$.
Here $\mathbb{F}_2^t = R^{s}/\im H_X$, and \eqref{eq: free res chain} is a free (hence projective) resolution of $\mathbb{F}_2^t$.
The result is
\begin{equation}
    (R/I)^{r} \xrightarrow{(H_Z^\dagger)_\ast} (R/I)^{q} \xrightarrow{(H_X)_\ast}  (R/I)^{s} \rightarrow \mathbb{F}_2^t \rightarrow 0,
\end{equation}
and the logical space is given by the $\Tor$ functor:
\begin{equation}
\ker (H_X)_\ast/\im (H_Z^\dagger)_\ast = \Tor_1^R(\mathbb{F}_2^t, R/I) = \left(\frac{(x-1,y-1)\cap I}{(x-1,y-1)I}\right)^{\oplus t} \cong \mathbb{F}_2^{2t}.
\end{equation}
The quotient $(x-1,y-1)\cap I / (x-1,y-1)I$ can be identified with the $\mathbb{F}_2$-vector space with basis $\{x^a-1,y^b-1\}$.
As expected, the number of logical qubits matches that of $t$ toric-code sectors.
Together with the full characterization of unbroken translation symmetry for anyons, this generalizes the twisted-boundary-condition results for bivariate bicycle codes in Ref.~\cite{liangGeneralizedToricCodes2025a} to arbitrary 2D translation-invariant topological CSS codes.

To construct decoupling maps for codes with PBC, observe that tensoring preserves chain isomorphisms.
The chain isomorphism \eqref{eq: chain map} from Theorem~\ref{thm: main} therefore remains a valid decoupling map after tensoring with $R/I$.
Moreover, a chain isomorphism between projective resolutions induces a canonical isomorphism between the corresponding left derived functors, ensuring that the logical operators of the two codes are identified correctly.
Concretely, tensoring \eqref{eq: chain map} with $R/I$ yields the following commutative diagram.
\begin{equation}
\begin{tikzcd}
	{(R/I)^{r}} & {(R/I)^{q}} & {(R/I)^{s}} & \mathbb{F}_2^t & 0 \\
	{(R/I)^{r}} & {(R/I)^{q}} & {(R/I)^{s}} & \mathbb{F}_2^t & 0
	\arrow["{(H_Z^\dagger)_\ast}", from=1-1, to=1-2]
	\arrow["{(H_X)_\ast}", from=1-2, to=1-3]
	\arrow[from=1-3, to=1-4]
	\arrow[from=1-4, to=1-5]
	\arrow["\psi_2\otimes_R \id", from=1-1, to=2-1]
	\arrow["{(\tilde{H}_Z^\dagger)_\ast}"', from=2-1, to=2-2]
	\arrow["\psi_1\otimes_R \id", from=1-2, to=2-2]
	\arrow["{(\tilde{H}_X)_\ast}"', from=2-2, to=2-3]
	\arrow["{\psi_0\otimes_R\id}", from=1-3, to=2-3]
	\arrow["{}", from=1-4, to=2-4]
	\arrow[from=2-3, to=2-4]
	\arrow[from=2-4, to=2-5]
\end{tikzcd}
\label{eq: chain map pbc}
\end{equation}
As maps on codes, $\psi_1\otimes_R \id$, $\psi_2\otimes_R \id$, and $\psi_0\otimes_R \id$ act locally in the same way as $\psi_1$, $\psi_2$, and $\psi_0$ on the infinite plane.

It is interesting to consider the special case $a=b=1$.
Now $R/I = \mathbb{F}_2$, and tensoring with $R/I$ amounts to evaluating at $x=y=1$.
In this case $\phi_1$ represents a decoding circuit
After applying the circuit represented by $\phi_1$, the code is decoupled into product states and $2t$ physical qubits, each carrying one logical qubit of the code.

\section{Locality and Time Complexity}\label{sec: linearsystem}

We now estimate the time complexity of Algorithm~\ref{alg: constructing isomorphism} and the locality of the resulting decoupling map.
To quantify locality, we use Laurent degree as a coarse measure of support size.
For a Laurent monomial $x^ay^b$, define $\deg(x^ay^b)=|a|+|b|$, and extend $\deg$ to polynomials, vectors, matrices, and finite sets by taking the maximum degree of a nonzero entry or element.
An entry of Laurent degree at most $D$ only couples unit cells within $\ell_1$ distance $D$ on the plane.
Thus the degree bounds below also bound the locality radius of the corresponding translation-invariant map.

The expensive part of Algorithm~\ref{alg: constructing isomorphism} is solving three Laurent-polynomial linear systems:
\begin{equation}
\begin{aligned}
H_X\phi_1 &= \tilde H_X, &
H_Z^\dagger\phi_2 &= \phi_1\tilde H_Z^\dagger, &
(\tilde H_Z^\dagger)^\top c^\top &= (\phi_2'-\phi_2)^\top .
\end{aligned}
\label{eq:algorithm-three-solves}
\end{equation}
The initial $\mathbb{F}_2$-linear change of basis costs only $O(q^3)$ operations and does not affect Laurent degrees.

We therefore analyze a general solvable system $AX=Y$ over $R=\mathbb{F}_2[x^{\pm1},y^{\pm1}]$, with $A\in R^{m\times n}$ and $Y\in R^{m\times k}$.
The method is Gr\"obner-basis reduction after lifting to the ordinary polynomial ring $P=\mathbb{F}_2[x,u,y,v]$, where $R\cong P/(xu-1,yv-1)$.
This lift is necessary because standard Buchberger-style Gr\"obner-basis algorithms use a well-ordered monomial set, while Laurent monomials allow arbitrarily negative exponents.
Concretely, we lift the columns of $A$ to $P^m$ and add the quotient-relation generators $(xu-1)e_i$ and $(yv-1)e_i$ for $1\le i\le m$.
The lift and the final descent modulo $(xu-1,yv-1)$ are lower-order preprocessing and postprocessing steps compared with the Gr\"obner-basis computation; moreover, the descent does not increase Laurent degree.

Let $\delta_A=\max\{\deg(A),2\}$ and $\delta_Y=\deg(Y)$, where the constant $2$ accounts for the generators $xu-1$ and $yv-1$.
After lifting, the module in $P^m$ is generated by the $n$ lifted columns of $A$ together with the $2m$ quotient-relation generators, so the number of generators is $n+2m$ and their degrees are bounded by $\delta_A$.
Liang's non-graded module Gr\"obner-basis bound~\cite{liang2022degree}, specialized to four polynomial variables and a free module of rank $m$, gives the degree cutoff $L_{\mathrm{GB}}=2(\delta_A m)^{16}$ for a reduced Gr\"obner basis.
A degree-by-degree Macaulay-matrix computation~\cite{faugereNewEfficientAlgorithm1999,BARDET201549} therefore only needs module monomials up to degree $L_{\mathrm{GB}}$.
Since there are $O((n+2m)L_{\mathrm{GB}}^4)$ such lifted rows and columns, cubic elimination gives the Gr\"obner-basis cost $O((n+2m)^3L_{\mathrm{GB}}^{12})=O((n+2m)^3(\delta_A m)^{192})$.

We also track the row reductions.
Let $f_i$ be the original lifted generators, namely the lifted columns of $A$ and the quotient-relation generators, and let $g_j$ be the reduced Gr\"obner-basis elements.
Tracking records $g_j=\sum_i v_{j,i}f_i$ with $\deg v_{j,i}\le L_{\mathrm{GB}}$.
For one lifted right-hand side column $y$, solvability of $AX=Y$ means that reduction by this basis has zero remainder, so $y=\sum_j u_jg_j$.
Degree-compatible reduction gives $\deg u_j\le\delta_Y$.
The tracked expression then recovers the solution coefficients from the components attached to the first $n$ lifted columns of $A$; the remaining $2m$ components multiply the quotient-relation generators and vanish after descending to $R$.
Thus each recovered coefficient is a sum of terms $u_jv_{j,i}$ with $\deg(u_jv_{j,i})\le\delta_Y+L_{\mathrm{GB}}$, so $\deg(X)\le\delta_Y+L_{\mathrm{GB}}=\delta_Y+2(\delta_A m)^{16}$.
For fixed or small $k$, the cost of reducing the right-hand sides and reading off the tracked coefficients is lower order than the Gr\"obner-basis computation, so we keep only the dominant term in the displayed time estimate:
\begin{equation}
T=O\!\left((n+2m)^3(\delta_A m)^{192}\right),\qquad \deg(X) \le \delta_Y+2(\delta_A m)^{16}.
\label{eq:laurent-linear-solve-bound}
\end{equation}

Now set $d_0=\max\{2,\deg(H_X),\deg(H_Z^\dagger)\}$ and apply Eq.~\eqref{eq:laurent-linear-solve-bound} to the three systems in Eq.~\eqref{eq:algorithm-three-solves}.
The relevant parameters and resulting bounds are:
\begingroup
\scriptsize
\begin{equation}
\begin{array}{c|c|c|c|c|c}
\text{system} & \text{unknown} & (m,n,k) & \delta_A & \delta_Y & \text{bound} \\
\hline
H_X\phi_1=\tilde H_X & \phi_1 & (s,q,q) & d_0 & 1 & \deg(\phi_1)=O(d_0^{16}q^{16}),\ T=O(d_0^{192}q^{195}) \\
H_Z^\dagger\phi_2=\phi_1\tilde H_Z^\dagger & \phi_2 & (q,r,r) & d_0 & O(d_0^{16}q^{16}) & \deg(\phi_2)=O(d_0^{16}q^{16}),\ T=O(d_0^{192}q^{195}) \\
(\tilde H_Z^\dagger)^\top c^\top=(\phi_2'-\phi_2)^\top & c & (r,q,r) & 2 & O(d_0^{16}q^{16}) & \deg(c)=O(d_0^{16}q^{16}),\ T=O(d_0^{192}q^{195})
\end{array}
\label{eq:algorithm-solve-bounds}
\end{equation}
\endgroup
The matrices $\xi_2$, $\xi_1$, and $\xi_0$ are degree-zero $\mathbb{F}_2$-matrices, so passing from $(\phi_2',\phi_1')$ to $(\xi_2\phi_2',\xi_1 \phi_1',\xi_0)$ does not change the degree bounds.
Thus Algorithm~\ref{alg: constructing isomorphism} returns the chain maps $(\xi_2\phi_2',\xi_1 \phi_1',\xi_0)$ with
\begin{equation}
T_{\mathrm{alg}}=O\!\left(d_0^{192}q^{195}\right),\qquad
\deg(\phi_2')=O\!\left(d_0^{16}q^{16}\right),\qquad
\deg(\phi_1')=O\!\left(d_0^{16}q^{16}\right).
\label{eq:algo-locality-time}
\end{equation}
A similar analysis gives polynomial time and degree bounds
for the inverses of these chain isomorphisms.

The worst-case exponents in Eq.~\eqref{eq:algo-locality-time} are intentionally conservative. As reported in Sec.~\ref{sec:bb-more-results}, the sampled benchmark and additional BB-code instances exhibit substantially milder scaling, with $T_{\mathrm{compute}} \propto q^{1.86}$ and $\deg(\psi_1^{-1})\propto q^{0.65}$ for $q=2\Lambda$.

\section{Applications}\label{sec: applications}
We close by describing two applications of the decoupling map: decoding by transporting syndromes and corrections through the map, and studying logical gates through the induced identification with the toric-code sectors.

\subsection{A unitary-decouple-based decoder}
\label{sec:unitary-decouple-based-decoder}

In this section, we introduce the \emph{unitary-decouple-based decoder}. Based on Theorem~\ref{thm: main} and Algorithm~\ref{alg: constructing isomorphism}, we find the explicit decoupling that maps a coarse-grained 2D translation-invariant topological code to stacks of toric codes and product states.
Decoding then proceeds in three steps:
\begin{enumerate}
    \item Map the measured syndrome of the original code to the TC sectors;
    \item Decode each TC sector independently using a graph-matching routine such as MWPM;
    \item Lift the direct sum of the sector corrections back to a Pauli operator on the original code.
\end{enumerate}

\begin{figure}[htbp]
    \centering
    \includegraphics[width=0.95\textwidth]{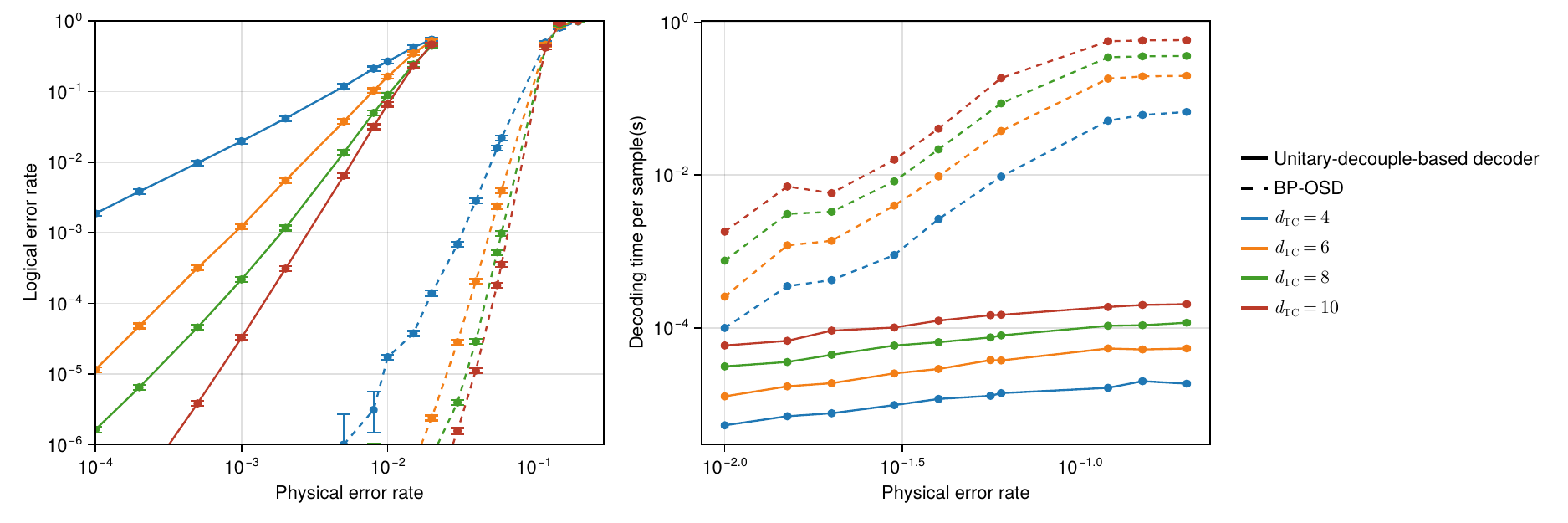}
    \caption{Numerical results for the unitary-decouple-based decoder on the BB code family defined by $H_X=(1+x+\overline{x}y,1+y+xy)$ and $H_Z=(1+\overline{y}+\overline{x}\overline{y},1+\overline{x}+x\overline{y})$. 
    Each instance is obtained by placing the superlattice on a torus with $d_{\mathrm{TC}} \times d_{\mathrm{TC}}$ squares.
    The left panel shows logical error rate versus physical error rate, while the right panel shows decoding time per sample.}
    \label{fig:unitary-decoupling-decoder-result}
\end{figure}

However, the performance of this decoder is not entirely satisfactory. Because the decoupling is implemented by a constant-depth circuit, a single local physical error in the original code can spread to several correlated qubits across multiple TC sectors. The independent matching subroutines ignore these inter-sector correlations, leading to a suboptimal global correction. We demonstrate this behavior in the numerical example below. In Fig.~\ref{fig:unitary-decoupling-decoder-result}, we compare the performance of the unitary-decouple-based decoder with BP-OSD~\cite{Roffe_2020} on BB codes defined by the check matrices:
\begin{equation}
    H_X = \begin{pmatrix}
        1+x+\overline{x}y & 1+y+xy
    \end{pmatrix},\quad
    H_Z = \begin{pmatrix}
        1+\overline{y}+\overline{x}\overline{y} & 1+\overline{x}+x\overline{y}
    \end{pmatrix}.
\end{equation}
For the numerical simulations, we use depolarizing errors under the code-capacity noise model.

As illustrated in Figure~\ref{fig:unitary-decoupling-decoder-result}, the unitary-decouple-based decoder exhibits a higher logical error rate compared to BP-OSD, yet it achieves a significantly faster decoding time. This empirical observation illustrates the speed-accuracy tradeoff inherited from the independent MWPM approximations.

Recently, several matching-based decoders have been proposed for BB codes, most notably the \emph{cell-matching decoder}~\cite{tan2026generalized} and the symmetry-based decoder (\emph{symatch})~\cite{sahay2026matching}. 
Closely related to our decoupling approach is the cell-matching decoder. It exploits the effective toric code decomposition by \emph{flushing}: projecting the syndrome onto $\coker H_{X/Z}$.
In contrast, the symatch decoder takes a more global route. It relies on \emph{code symmetries} and the \emph{cylinder trick} to extract the commutation data of logical operators.

\subsection{Implication on logical gates}

In the main text, we have discussed some potential applications of our results
to constructing logical gates for topological codes.
Here we review the example of the color codes, which establishes the macroscopic picture of the relationship between logical gates and the action on anyon strings from different toric-code sectors.
We then present the results of a numerical search on transporting transversal CNOT gates from toric codes to general topological CSS codes.
We focus on the operator spreading behavior for such constructions.

\paragraph{Color-code example.}
Consider a two-dimensional color code on a torus,
which is locally Clifford equivalent to two toric-code sectors
and therefore encodes four logical qubits~\cite{kubicaUnfoldingColorCode2015}.
Let $h$ and $v$ denote two noncontractible cycles intersecting once,
and let $A$ and $B$ denote two independent color-code string colors.
Here $X_{c,\gamma}$ and $Z_{c,\gamma}$ denote color-$c$ string operators
supported along $\gamma$.
Strings of different colors on the two intersecting cycles provide
canonical logical Pauli pairs~\cite{bombinStatisticalMechanicalModels2008}.
Let $S_{\mathrm{CC}}$ denote the transversal phase operation.
For the 6.6.6 color code, this operation applies $S$ and $S^\dagger$
on the two vertex superlattices; up to Pauli phases,
$X\mapsto XZ$ and $Z\mapsto Z$.
Choosing the logical basis shown below, its complete action is
given by the following array, where a prime denotes conjugation by
$S_{\mathrm{CC}}$:
\begingroup
\renewcommand{\arraystretch}{1.15}
\setlength{\arraycolsep}{3.5pt}
\small
\[
    \begin{array}{cccc@{\qquad}cccc}
        \hline
        \bar X_i & X_{c,\gamma} & X'_{c,\gamma} & \bar X_i'
        & \bar Z_i & Z_{c,\gamma} & Z'_{c,\gamma} & \bar Z_i' \\
        \hline
        \bar X_1 & X_{A,v} & X_{A,v}Z_{A,v} & \bar X_1\bar Z_4
        & \bar Z_1 & Z_{B,h} & Z_{B,h} & \bar Z_1 \\
        \bar X_2 & X_{B,v} & X_{B,v}Z_{B,v} & \bar X_2\bar Z_3
        & \bar Z_2 & Z_{A,h} & Z_{A,h} & \bar Z_2 \\
        \bar X_3 & X_{A,h} & X_{A,h}Z_{A,h} & \bar X_3\bar Z_2
        & \bar Z_3 & Z_{B,v} & Z_{B,v} & \bar Z_3 \\
        \bar X_4 & X_{B,h} & X_{B,h}Z_{B,h} & \bar X_4\bar Z_1
        & \bar Z_4 & Z_{A,v} & Z_{A,v} & \bar Z_4 \\
        \hline
    \end{array}
\]
\endgroup
For instance, $\bar X_1=X_{A,v}$ acquires the same-color,
same-support string $Z_{A,v}=\bar Z_4$.
The table therefore gives
$\bar S_{\mathrm{CC}}
=\overline{\mathrm{CZ}}_{1,4}\overline{\mathrm{CZ}}_{2,3}$.
Under unfolding, logical qubits $\{1,3\}$ and $\{2,4\}$ form the two
toric-code sectors, respectively, so both controlled-$Z$ gates couple
the two toric-code sectors.

\begin{figure}
    \centering
    \includegraphics[width=.8\textwidth]{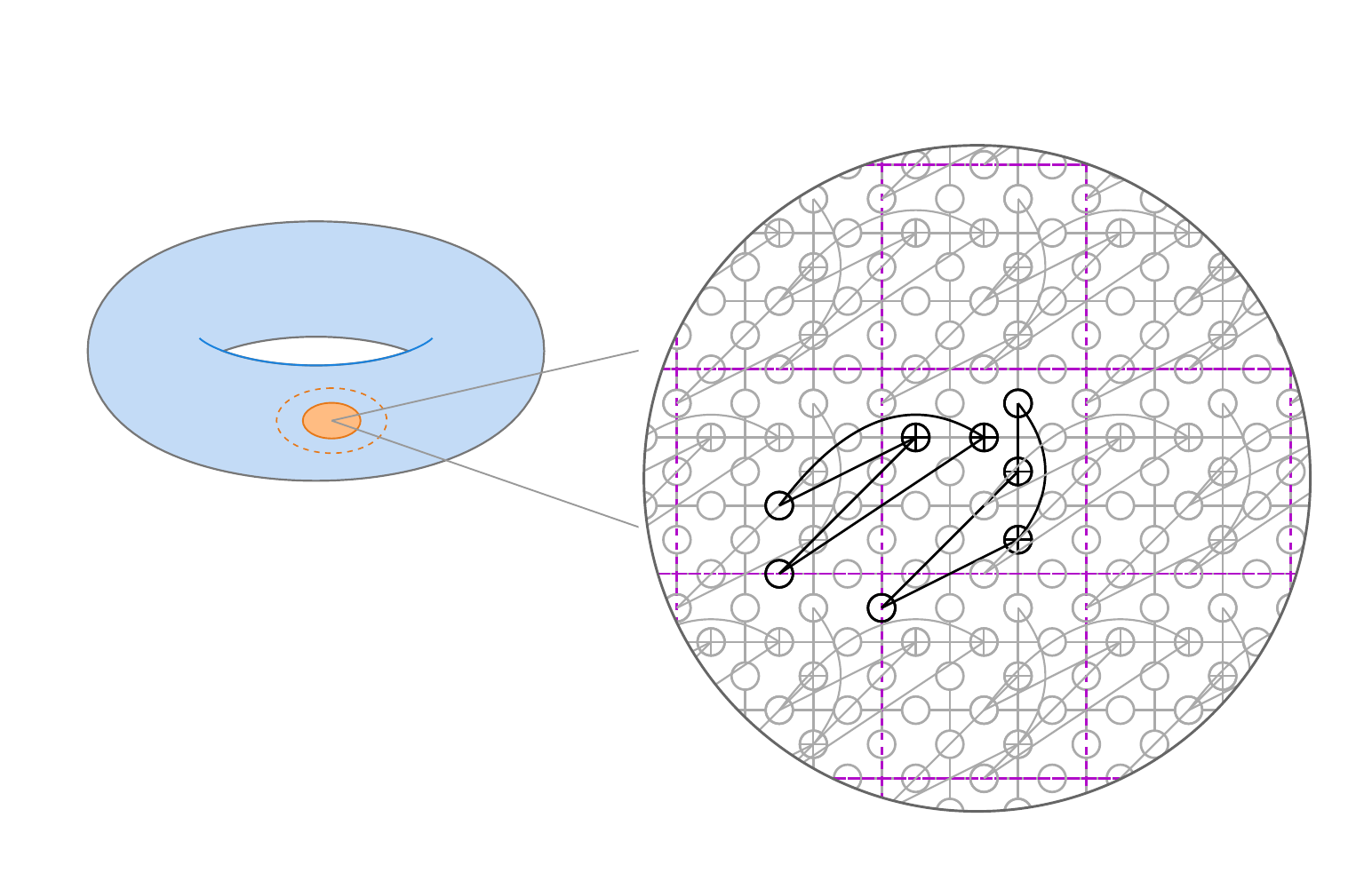}
    \caption{
    Realization of the transported CNOT $1\to2$ for the
    BB code $H_X=(1+x+\bar{x},1+y+\bar{y})$.  The pattern contains
    eight CNOT couplings per coarse-grained unit cell and has
    column-support spreading $s(Q_{1\to2})=3$.
    }
    \label{fig:cnot-demo}
\end{figure}

\paragraph{Transporting inter-copy CNOTs.}
The decoupling map can also be used in the constructive direction.
Let $E_{j\to i}$ denote the qubit-wise CNOT from toric-code sector $j$
to sector $i$ in the standard complex.
In the convention used in our numerical calculation, its action pulled
back to the original code is
\begin{equation}
    Q_{j\to i}=\phi_1 E_{j\to i}\phi_1^{-1}.
    \label{eq:transported-cnot}
\end{equation}
As an invertible translation-invariant CSS map, $Q_{j\to i}$ can be
compiled using the procedure in Sec.~\ref{sec: removing shifts}.
Here we compute its operator spreading.
We quantify the support spreading of this map by
$s(Q)=\max_k\sum_l|\operatorname{supp} Q_{lk}|$, the largest number of Laurent
monomials in any column.  Thus $s(Q)$ measures the largest support of
the image of a basis Pauli in this CSS sector, rather than the number
of CNOT gates in a circuit decomposition.

Table~\ref{tab:transported-cnot-summary} lists all ordered pairs of
toric-code sectors for six representative BB codes.  The construction
exists for every pair, but its spreading depends strongly on both the
code and the direction of the CNOT.
\begin{table*}[htbp]
    \centering
    \begingroup
    \scriptsize
    \setlength{\tabcolsep}{2.6pt}
    \renewcommand{\arraystretch}{1.04}
    \begin{tabular*}{\textwidth}{@{\extracolsep{\fill}}ccccc@{\qquad}ccccc}
        \hline
        $f$ & $g$ & Control $j$ & Target $i$ & $s(Q_{j\to i})$
        & $f$ & $g$ & Control $j$ & Target $i$ & $s(Q_{j\to i})$ \\
        \hline
        \multirow{2}{*}{$1+x+xy$}
        & \multirow{2}{*}{$1+y+xy$}
        & 2 & 1 & 25
        & \multirow{2}{*}{$1+x+x^2$}
        & \multirow{2}{*}{$1+y+x^2$}
        & 2 & 1 & 9 \\
        & & 1 & 2 & 7
        & & & 1 & 2 & 9 \\
        \hline
        \multirow{6}{*}{$1+x+x^{-1}y$}
        & \multirow{6}{*}{$1+y+xy$}
        & 2 & 1 & 15
        & \multirow{12}{*}{$1+x+x^{-1}$}
        & \multirow{12}{*}{$1+y+y^{-1}$}
        & 2 & 1 & 7 \\
        & & 3 & 1 & 15
        & & & 3 & 1 & 21 \\
        & & 1 & 2 & 15
        & & & 4 & 1 & 17 \\
        & & 3 & 2 & 21
        & & & 1 & 2 & 3 \\
        & & 1 & 3 & 11
        & & & 3 & 2 & 11 \\
        & & 2 & 3 & 11
        & & & 4 & 2 & 11 \\
        \cline{1-5}
        \multirow{12}{*}{$1+x+x^{-1}$}
        & \multirow{12}{*}{$1+y+x^3y^2$}
        & 2 & 1 & 15
        & & & 1 & 3 & 7 \\
        & & 3 & 1 & 17
        & & & 2 & 3 & 7 \\
        & & 4 & 1 & 21
        & & & 4 & 3 & 23 \\
        & & 1 & 2 & 39
        & & & 1 & 4 & 7 \\
        & & 3 & 2 & 17
        & & & 2 & 4 & 7 \\
        & & 4 & 2 & 21
        & & & 3 & 4 & 21 \\
        \cline{6-10}
        & & 1 & 3 & 61
        & \multirow{6}{*}{$1+x+xy$}
        & \multirow{6}{*}{$1+y+xy^{-1}$}
        & 2 & 1 & 15 \\
        & & 2 & 3 & 17
        & & & 3 & 1 & 17 \\
        & & 4 & 3 & 29
        & & & 1 & 2 & 39 \\
        & & 1 & 4 & 57
        & & & 3 & 2 & 39 \\
        & & 2 & 4 & 21
        & & & 1 & 3 & 43 \\
        & & 3 & 4 & 23
        & & & 2 & 3 & 39 \\
        \hline
    \end{tabular*}
    \caption{Support spreading of transported inter-copy CNOTs.
    For each BB code, all $t(t-1)$ ordered pairs of its toric-code
    sectors are shown.}
    \label{tab:transported-cnot-summary}
    \endgroup
\end{table*}

\begin{example}
    Consider the BB code
    $H_X=(1+x+\bar{x},1+y+\bar{y})$.
    For the pair $1\to2$, the transported map has $s(Q_{1\to2})=3$.
    Figure~\ref{fig:cnot-demo} shows a periodic realization using eight
    CNOT couplings per coarse-grained unit cell.  Qubit-wise CNOT
    between the two toric-code sectors implements CNOTs between their
    corresponding encoded qubits~\cite{gottesmanStabilizerCodesQuantum1997}.
\end{example}

\bibliography{main}